\documentclass[journal ]{new-aiaa}
\usepackage[utf8]{inputenc}
\usepackage{textcomp}

\usepackage{graphicx}
\usepackage{amsmath}
\usepackage[version=4]{mhchem}
\usepackage{siunitx}
\usepackage{longtable,tabularx}
\usepackage{subfigure}
\usepackage{float}
\usepackage[table]{xcolor}
\usepackage[justification=centering]{caption}
\usepackage[most]{tcolorbox}
\usepackage{bm}
\usepackage{bbm}

\renewcommand{\vec}[1]{\bm{#1}}

\title{Information-Based Sensor Placement for Data-Driven Estimation of Unsteady Flows}

\author{John Graff\footnote{Graduate Student, Department of Mechanical and Aerospace Engineering. AIAA Student Member.}}
\affil{University at Buffalo, The State University of New York, Buffalo, NY, 14260}
\author{Albert Medina\footnote{Research Scientist, Aerodynamic Technology Branch, Aerospace Systems Directorate. AIAA Senior Member.}}
\affil{Air Force Research Laboratory, Wright-Patterson Air Force Base, Ohio, 45433}
\author{Francis D. Lagor\footnote{Assistant Professor, Department of Mechanical and Aerospace Engineering. AIAA Senior Member.}}
\affil{University at Buffalo, The State University of New York, Buffalo, NY, 14260}

\begin{document}

\maketitle

\begin{abstract}
Estimation of unsteady flow fields around flight vehicles may improve flow interactions and lead to enhanced vehicle performance. Although flow-field representations can be very high-dimensional, their dynamics can have low-order representations and may be estimated using a few, appropriately placed measurements.  This paper presents a sensor-selection framework for the intended application of data-driven, flow-field estimation. This framework combines data-driven modeling, steady-state Kalman Filter design, and a sparsification technique for sequential selection of sensors. This paper also uses the sensor selection framework to design sensor arrays that can perform well across a variety of operating conditions.  Flow estimation results on numerical data show that the proposed framework produces arrays that are highly effective at flow-field estimation for the flow behind and an airfoil at a high angle of attack using embedded pressure sensors.  Analysis of the flow fields reveals that paths of impinging stagnation points along the airfoil's surface during a shedding period of the flow are highly informative locations for placement of pressure sensors.
\end{abstract}

\section*{Nomenclature}

{\renewcommand\arraystretch{1.0}
\noindent\begin{longtable*}{@{}l @{\quad=\quad} l@{}}
$\hat{A}$ & matrix of inner products of feature vectors $\vec{\xi}(\vec{x}_i)$ and $\vec{\xi}(\vec{x}^{\prime}_j)$\\
$\vec{b}_k$ & vector of weights applied to $\Psi_r$ to produce $\vec{x}_k$\\
$\text{cov}(\cdot)$ & covariance calculation\\
$d$ & scaling factor in polynomial kernel\\
$F$ & dynamics matrix used in Koopman Observer Form\\
$f(\cdot , \cdot)$ & kernel function\\
$\hat{G}$ & matrix of inner products of feature vectors $\vec{\xi}(\vec{x}_i)$ and $\vec{\xi}(\vec{x}_j)$\\
$\vec{g}(\vec{x})$ & set of observable functions with linear dynamics\\
$H_x$ & matrix that maps the modal state to the system state\\
$H_y$ & matrix maps the modal state to the system outputs\\
$I_\infty$ & steady-state information matrix ($P_\infty^{-1}$)\\
$\text{Im}(\cdot)$ & operator that extracts the imaginary component of a complex number\\
$i,j,k$ & indices\\
$K_k$ & Kalman gain matrix at time step $k$\\
$K_\infty$ & steady state Kalman gain matrix\\
$K_\text{DMD}$ & linear operator used in Dynamic Mode Decomposition\\
$K_\text{EDMD}$ & linear operator used in Extended Dynamic Mode Decomposition\\
$\hat{K}$ & approximation of the Koopman operator\\
$\mathcal{K}$ & Koopman operator\\
$N$ & number of snapshots\\
$n_c$ & number of candidate sensors\\
$n_s$ & number of selected sensors\\
$P_k$ & error covariance matrix for time step $k$\\
$P_\infty$ & steady-state error covariance matrix\\
$Q$ & process noise covariance matrix\\
$Q_x$ & matrix containing Koopman modes in its rows\\
$Q_y$ & matrix containing output Koopman modes in its rows\\
$R$ & measurement noise covariance matrix\\
$R_d$ & diagonal matrix with entries extracted from the diagonal of $R$\\
$\text{Re}(\cdot)$ & operator that extracts the real component of a complex number\\
$r$ & rank of truncation of reduced-order model\\
$V$ & matrix containing eigenvectors of $\hat{G}$\\
$\vec{v}_k$ & measurement noise with distribution ${\cal N}\left(\vec{0},R\right)$ at time step $k$\\
$\vec{w}_{k-1}$ & process noise with distribution ${\cal N}\left(\vec{0},Q\right)$ at time step $k-1$\\
$X$ & matrix containing $N$ snapshots of the state in its rows\\
$\bar{X}$ & matrix containing $N$ snapshots output appended onto the state in its rows\\
$\vec{x}_k$ & state vector at time instant $k$\\
$Y$ & matrix containing $N$ snapshots of the output in its rows\\
$\vec{y}_k$ & output vector at time instant $k$\\
$\vec{z}_k$ & modal-state vector with only real entries at time instant $k$\\

$\alpha$ & parameter that determines the order of the polynomial kernel\\
$\Theta$ & matrix containing the eigenvectors of $\hat{K}$\\
$\Lambda$ & diagonal matrix containing the eigenvalues of $\hat{K}$\\
$\vec{\mu}(\vec{\omega})$ & a continuous, nondecreasing, concave map of the resource allocation weights, $\vec{\omega}$\\
$\vec{\xi}(\vec{x})$ & dictionary of observable functions chosen for use in EDMD and KDMD\\
$\Sigma$ & diagonal matrix containing the square root of the eigenvalues of $\hat{G}$\\
$\Phi(\cdot)$ & function that returns a scalar measure of the value of a matrix\\
$\Phi_x$ & matrix containing Koopman eigenfunction values\\
$\Psi_r$ & matrix containing the $r$ most energetic POD modes of $X^T$\\
$\bar{\Psi}$ & matrix containing the POD modes of $\bar{X}^T$\\
$\bar{\Psi}_y$ & matrix containing only the output portions of the $r$ most energetic POD modes of $\bar{X}^T$\\
$\vec{\omega}$ & resource allocation weights\\

$(\cdot)^{\prime}$ & offset of one or more timesteps\\
$(\cdot)^\dagger$ & Moore-Penrose pseudo-inverse\\
$\hat{(\cdot)}$ & estimated or approximated quantity\\
$(\cdot)^f$ & forecasted quantity that does not include assimilation of data at the current time step\\
$(\cdot)^a$ & quantity that includes assimilation of data available at the current time step
\end{longtable*}}

%%%%%%%%%%%%%%%%%%%%%%%%%%%%%%%%%%%%%%%%%%%%%%%%%%%%%%%%%%%%%%%%%%%%%%%%%%%%%%%%%%%%%%%%%%%%%%%%%%%%%%%%%%%%%%%%%%%%

\section{Introduction}
\lettrine{F}low-field modeling and estimation are active areas of research \cite{Tu_2013,Gomez_2019,Graff_2019,Ahuja_2010,Rowley_Dawson_2017}, because better understanding of the unsteady flow around a flight vehicle may enhance performance through gains in stability and control.  This paper focuses on the specific task of unsteady flow estimation over an aircraft surface using the surface pressure measurements.  Given practical constraints that are often present in the design of flight vehicles, such as cost and space limitations, careful placement of sparse sensors is important to make best use of measurement resources.  Hence, sensor placement is a crucial component in the viability and accuracy of flow estimation.

Due to potential sensing gains, the question of optimal sensor placement has been widely researched.  However, it remains a challenging and ongoing research area.  To select $n_s$ sensors from a set of $n_c$ candidate sensors to optimize an appropriate performance measure, the brute-force optimal solution requires search across all possible combinations of $n_s$ sensors chosen from $n_c$ candidates.  The combinatorial growth of the computational cost often makes the brute-force optimal solution unattainable, even for relatively small values of $n_s$ and $n_c$.  In 2008, Joshi and Boyd \cite{Joshi_2008} developed a sensor selection method for problems involving linear sensors with additive measurement noise that uses semi-definite programming to minimize of the log determinant of the error covariance matrix. Their selection method reduces the computational complexity of array design to $\mathcal{O}(n_s^2)$, but there is no guarantee that there is only a small gap between the performance of the chosen sensor array and the optimal performance bound.  In 2010, Shamaiah et al. \cite{Shamaiah_2010} proposed a greedy algorithm for sensor selection that guaranteed its selection to be within $(1- 1/e)$ of the optimal solution while further reducing the computational cost.  In 2020, Hashemi et al. \cite{Hashemi_2020} proposed a randomized and greedy sensor-selection algorithm that further reduced computational cost with the objective of minimizing the mean-square error of the Kalman filter.  In the current paper, the computational cost is addressed by solving a convex relaxation of the optimal sensor placement problem using semidefinite programming, similar to \cite{Joshi_2008}.  

Another important reduction in the computational cost of sensor placement derives from use of a reduced-order flow-field model, since model evaluation is required for each assessment of a candidate configuration.  Estimation of the high-dimensional state of a flow field can be achieved using few measurements if the underlying dynamics of the flow field are low-dimensional.   For a system with low-dimensional dynamics, the majority of the flow's energy can be represented using only a few modes of the Proper Orthogonal Decomposition (POD).  POD is a long-established, reduced-order modeling tool that makes use of the Singular Value Decomposition (SVD) of the data to create modes that that can represent the flow field when recombined in weighted combinations.  In 2006, Willcox \cite{WILLCOX2006} proposed use of Gappy POD to select pressure sensor locations on the surface of an airfoil such that the condition number of the correlation matrix formed from the POD basis vectors is minimized. This method assumes that the POD space has been truncated to produce a reduced-order representation of the system and then sensors are selected to preserve orthogonality of the POD basis vectors. 

Several other sensor placement methods based on POD have been presented in the literature, including \cite{Yang_2010,QR_Manohar,QR_Clark,Saito_2021}. Yang et al.\ \cite{Yang_2010} proposed placing sensors at the extrema of POD modes.  Manohar et al.\ \cite{QR_Manohar} placed sensors at the pivot locations resulting from the QR decomposition of a tailored basis for the system such as the POD modes for the case where the number of sensors is equal to the number of modes. They also extended this concept to the case where the number of sensors exceeds the number of POD modes, which is known as the oversampled case.  Clark et al.\ \cite{QR_Clark} further extended QR pivoting for sensor selection to the cost-constrained case where pivots are chosen not only to minimize error but also to minimize an associated cost of the pivot to be selected.  Saito et al.\ \cite{Saito_2021} introduced a determinant-based, greedy, sensor-selection algorithm and showed that it is mathematically equivalent to the QR-pivoting method from \cite{QR_Manohar} under certain conditions and more computationally efficient in the oversampled case where the number of sensors exceeds the number of POD modes.  In the current paper, POD plays an important role in model-reduction by providing a subspace in which the dynamics of the flow model can be conveniently represented.  

Many sensor placement methods assume that the candidate sensors directly (or linearly) measure components of the target of inference or reconstruction.  For example, sensor placement using Gappy Proper Orthogonal Decomposition \cite{WILLCOX2006}, QR-pivoting sensor placement \cite{QR_Manohar,QR_Clark}, convex-optimization-based sensor placement \cite{Joshi_2008}, and determinant-based fast greedy sensor selection sensor placement \cite{Saito_2021} all restrict their sensors to direct state measurements.  Such restrictions are often useful for analysis in sensor placement, because inference from sensor measurements is then confined to examination of (pseudo-)invertibility of the measurement equation and rank arguments.  However, direct inversion of the sensor measurements for state inference produces instantaneous estimates that are not based on prior measurements or prior estimates.  Instantaneous estimation may be suitable some applications, however, systems that have rich dynamics offer additional opportunity for improved localization of the state.  Applying a dynamic estimator to such a system is useful because the time evolution of a measurement may provide sufficient information to determine the state, even if a measurement from a single time instant cannot.  Whether the state can be inferred from accumulation of output measurements in time is a question of observability in control theory, and observability measures have previously been used for sensor placement (e.g., see \cite{Lagor2013b,Hinson_2014,Lagor2016b,Bopardikar_2019}).  However, sensor placement that is based solely on observability does not optimize the sensing array for filtering applications due to its lack of prior information, as well as its lack of process and measurement noise.

Sensor placement methods have also been developed for dynamic estimators, such as the Kalman filter that is used in this paper.  Tzoumas et al.\ \cite{Tzoumas_2016} proposed two sensor-placement algorithms for a Kalman filter.  In the first algorithm, the log determinant of the Kalman filter error covariance matrix is set to a fixed value and the sparsest sensor array that produces the specified log determinant of the error covariance is selected.  The second algorithm sets a maximum number of sensors and minimizes the log determinant of the error covariance subject to this maximum array size.  In both algorithms, the array selection is dependent on the initial error covariance chosen.  Zhang et al.\ \cite{Zhang_2017} remove the dependence of the initial condition of the error covariance matrix by performing greedy sensor selection with the objective of minimizing the trace of the steady-state error covariance matrix for a steady-state Kalman filter.  The steady-state Kalman filter has an error covariance that satisfies an algebraic Riccati equation and is independent of the initial condition for a stable filter.  Zhang et al.\ \cite{Zhang_2017} show that their greedy approach performs optimally for systems with ordered information matrices.  In the current paper, the sensor placement framework for a Kalman filter that is proposed is similar to \cite{Zhang_2017} since it also applies to a steady-state Kalman filter.  Both \cite{Tzoumas_2016} and \cite{Zhang_2017} perform greedy selection that requires approximately $n_s n_c$ (i.e., number of sensors, times number of candidate sensors) evaluations of the objective function of the error covariance.  In constrast, the sequential approach of the current paper solves $n_s$ resource allocation calculations, which are solutions to a Semi-Definite Program (SDP) that correspond to a relaxation the discrete sensor-selection problem.  The proposed method can provide substantial computational savings in cases of a large number of candidate sensors or a time-consuming evaluation of the objective function of the error covariance.  

To obtain a model that is useful for flow estimation, this paper selects a data-driven modeling approach, so that sensor placement and estimation techniques can be readily applied to real systems.  The data-driven modeling techniques employed by the sensor placement framework presented in this paper are based on Koopman operator theory, which is a theoretical framework that advances observables of a system in time by providing a linear representation of the dynamics in an appropriate set of transformed coordinates \cite{Koopman_1931, Koopman_1932}.  Kernel-based Dynamic Mode Decomposition (KDMD) was developed by Williams et al. \cite{Williams_2015} as a method of approximating the Koopman operator for a system given a kernel that represents inner products of observable vectors for a chosen dictionary of observables. Surana and Banaszuk used KDMD to develop the Koopman Observer Form (KOF) \cite{Surana_2016}, which is a linear dynamic model of the system that is suitable for state estimation.  Gomez et al. \cite{Gomez_2019} applied the KOF with a Kalman filter to estimate the flow over an airfoil using embedded pressure measurements.  Following these works, this paper adopts the use of KDMD to produce a linear model of the unsteady fluid system that is in KOF. 

Sensor placement is performed using this model with an information-based resource-allocation method developed by Sagnol and Harmon \cite{Sagnol_2015}. They used semi-definite programming to optimally allocate sensor resources for a steady-state Kalman filter. The sensor selection method presented in \cite{Joshi_2008} assumes there is no process noise in the system's dynamics \cite{Tzoumas_2016}, which is a special case of the method presented in \cite{Sagnol_2015}. The semi-definite-programming  method from \cite{Sagnol_2015} optimally allocates resources to available sensors to maximize a user-defined scalarization of the information matrix, but there are no guarantees on the sparsity of the resulting resource allocation. The resulting resource allocation often does not match the desired sparsity, so we propose sequentially selecting the highest weighted sensor in the resource allocation and accounting for the influence of selected sensors on the information matrix. We propose two variants of our sparsification method: orthogonal selection, and complementary selection. Orthogonal selection chooses the highest weighted sensor and adjusts the output row vectors of the unselected sensors according to Gram-Schmidt orthogonalization. This ensures that the candidate sensors in the next selection step have output row vectors that are orthogonal to the output row vectors of previously selected sensors. Thereby, the resource-allocation solve in the next iteration weights sensors according to how their output row vectors help to maximize information measure beyond the contributions of the selected set. The second method is complementary selection. After each resource-allocation solve, complementary selection forces the weight of the selected sensor to 1 for all subsequent solves. This approach maximizes the information measure at each step but has no explicit requirements on the output row vectors of selected sensors. Hence, the sensors complement each other in maximizing the information measure, but the method does allow for correlated contributions from the selected output row vectors.

\begin{figure}[h!]
\centering
\includegraphics[width=.98\linewidth, clip=true, trim=0in 0in 0in 0in]{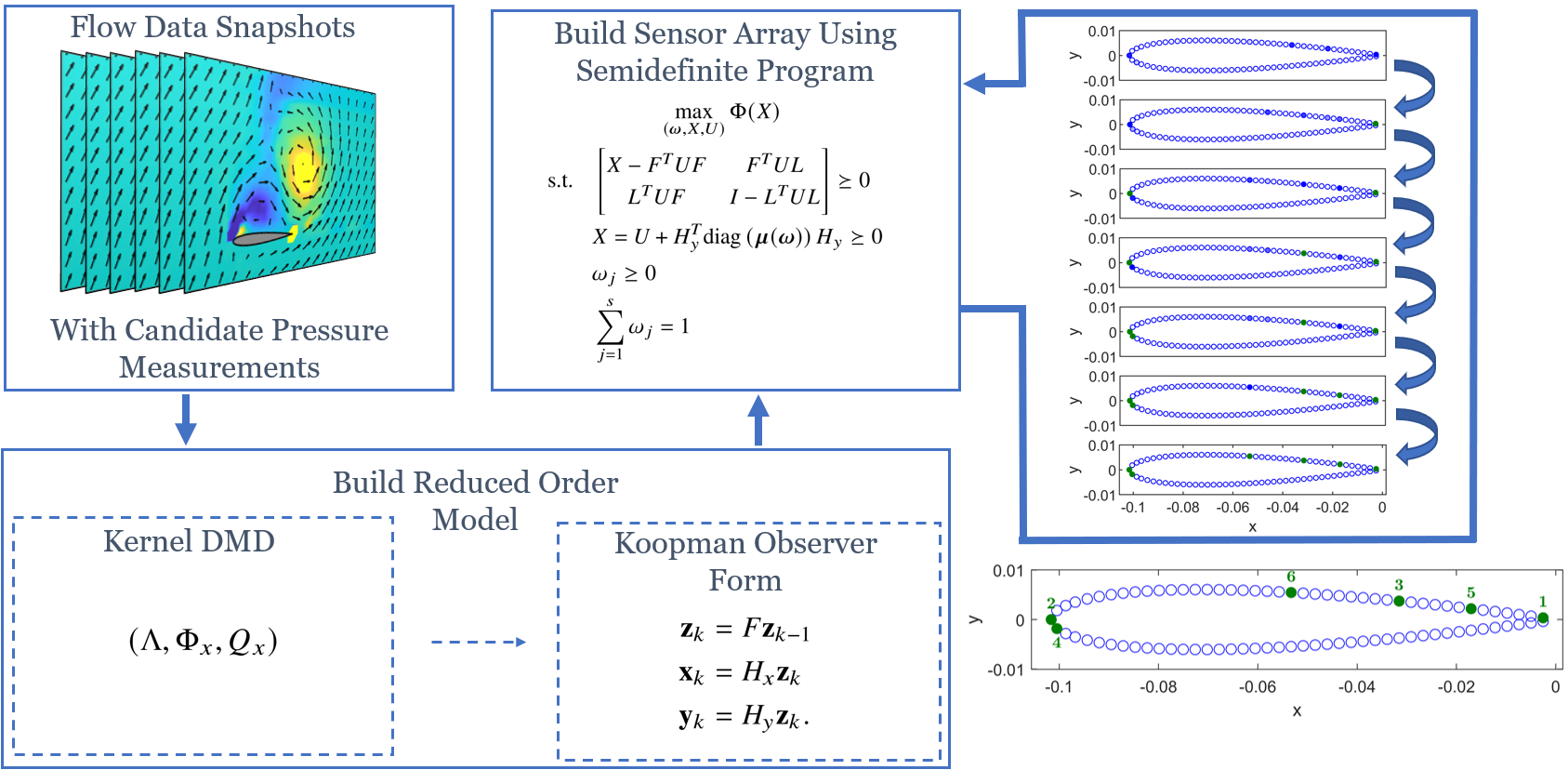}
\caption{Overview of proposed sensor selection process}
\label{fig:overview}
\end{figure}

Figure \ref{fig:overview} summarizes the framework that this paper contributes for data-driven sparse sensor selection for use in Kalman filtering.  The framework processes flow-field data using KDMD to build a linear of model of the fluid system that is put into KOF.  The system model is used within an SDP that solves for sensor allocations (i.e., weightings) that maximize a measure of the steady-state information matrix.  To make the sensor allocations sparse, sensors are sequentially selected.  Two variants of the selection process are proposed: orthogonal selection and complementary selection. We provide a comparison of the two proposed selection variants and a QR-based selection in a filtering problem of flow over an airfoil. This contribution delivers a step-by-step approach to obtain a flow model from data and use it to place sensors for flow estimation. The method is agnostic to sensor type, provided the sensor data is rich enough to yield a model meeting the observability condition for the steady-state Kalman filter. Filtering of simulated data for flow past an airfoil shows that the proposed framework performs well in its intended application.

Since data-driven methods can be sensitive to the operating condition, this paper also contributes a method to design sensor arrays that will perform flow estimation well across several operating conditions.  Through the construction of a composite system that examines the expected information measure for application of an array at all operating points, the sensor selection framework in Fig.\ \ref{fig:overview} is applied with appropriate weight constraints to build an array for all operating points simultaneously.  Numerical filtering experiments reveal that orthogonal and complementary selection both produce arrays that perform well for filtering across all operating points considered when the array is built on the composite system.  However, analysis of all arrays designed in this study over all operating conditions for information measure and filtering performance reveals that the choosing the array which maximizes the minimum information measure on to across design points is a good indicator of which array design will have the best average filtering performance across design points.  Lastly, this paper also contributes an analysis of informative sensor locations and their proximity to stagnation points of the flow field, which provide important topological information of large vortical structures in the flow.  Findings indicate that the most informative sensor locations occur near and along the paths taken by impinging stagnation points. 

The remainder of the paper is organized as follows. Section \ref{sec:Background} reviews tools necessary for data-driven flow estimation. Section \ref{sec:Sensor} presents our sparse sensor-selection framework. Section \ref{sec:Results} presents results from numerical experiments of pressure sensor placement and flow estimation at various operating conditions and examines how to design an array design across several design points. Section \ref{sec:flow_features} analyzes the location on sensors and nearby flow features.  Section \ref{sec:Conclusion} summarizes the paper and discusses future work.

%%%%%%%%%%%%%%%%%%%%%%%%%%%%%%%%%%%%%%%%%%%%%%%%%%%%%%%%%%%%%%%%%%%%%%%%%%%%%%%%%%%%%%%%%%%%%%%%%%%%%%%%%%%%%%%%%%%%

\section{Data-Driven Flow Estimation} \label{sec:Background}
This section provides a background on a technique for estimating a flow field from sensor measurements. The technique builds a data-driven model using Kernel Dynamic Mode Decomposition to approximate the Koopman Mode Decomposition.  Koopman Observer Form organizes the state and output quantities in the data-driven model, and a Kalman filter is used to estimate the state.

\subsection{Koopman Mode Decomposition}
For a discrete-time dynamical system \cite{Williams_2015}
\begin{equation}
\vec{x}_k = \vec{f}(\vec{x}_{k-1}),
\label{eqn:dynamics}
\end{equation}
let $\vec{g}(\vec{x})$ be a vector of observables that may contain directly observed components of the state vector $\vec{x}$ or functions of the state components. The Koopman operator $\mathcal{K}$ pushes scalar observables forward in time according to the dynamics \eqref{eqn:dynamics} such that \cite{Koopman_1931, Koopman_1932,Williams_2015}
\begin{equation}
\left(\mathcal{K}g_j\right)(\vec{x}) = g_j(\vec{f}(\vec{x}))
\end{equation}
for the $j$th observable component $g_j$.  Since the Koopman operator is a linear, infinite-dimensional operator, spectral analysis can help to study the properties of the operator. Let $\phi_j(\vec{x}_0)$ be the $j$th eigenfunction of Koopman operator $\mathcal{K}$ evaluated at the initial condition $\vec{x}_0$, and let $\lambda_j$ denote the corresponding eigenvalue. The time-evolution of a vector of observables can be expressed in terms of the Koopman eigenfunctions according to Koopman Mode Decomposition (KMD) \cite{Williams_2015}
\begin{equation}
\vec{g}(\vec{x}_k) = \sum_{j=1}^\infty \lambda_j^k\phi_j(\vec{x}_0)\vec{q}_j,
\label{eqn:KMD}
\end{equation}
where $\vec{q}_j$ is a vector of coefficients.  If the observable vector is a direct measurement of the state, such that $\vec{g}(\vec{x}_k) = \vec{x}_k$, then the coefficient vector $\vec{q}_j$ is referred to as a Koopman mode. Thereby, the Koopman eigenfunctions, eigenvalues, and modes are important for modeling the evolution of observables. Although the KMD can provide a linear model for evolution of an observable vector, the model may require an infinite number of Koopman eigenfunctions to fully achieve the linear representation. Since real applications require finite precision, the Koopman operator must be approximated in a suitable subspace of observables \cite{Williams_2015}.

Dynamic Mode Decomposition (DMD) is a method of data analysis that examines the best-fit, linear operator that marches a measurement snapshot forward by one time unit. A measurement snapshot at a given time is a vectorized collection of all measurements being studied. Consider a pair of snapshots $(\vec{x}_j,\vec{x}'_j)$ for $j= 1, \dots, N$, for which $\vec{x}_j$ and $\vec{x}'_j$ may be offset by one or more time units. To aid in the mathematical description of the DMD method, define the data matrices
\begin{equation}
X =
\begin{bmatrix}
\vec{x}_1^T \\
\vec{x}_2^T \\
\vdots \\
\vec{x}_N^T
\end{bmatrix}
\qquad\text{ and }\qquad
X' =
\begin{bmatrix}
\vec{x}^{\prime T}_1 \\
\vec{x}^{\prime T}_2 \\
\vdots \\
\vec{x}^{\prime T}_N
\end{bmatrix}.
\label{eqn:dataMats}
\end{equation}
DMD studies the operator $K_\text{DMD}$ that links the snapshots in the data matrices via the equation, $X' \approx X K_\text{DMD}$.

In DMD, the measured data are used as state components in \eqref{eqn:dataMats} directly; hence, the basis that DMD uses to represent the flow field contains only linear combinations of the state components \cite{Williams_2015}. A richer set of observables based on nonlinear functions of the state can provide a basis that is capable of representing eigenfunctions of the Koopman operator $\mathcal{K}$. This observation led to the development of Extended Dynamic Mode Decomposition (EDMD), which calculates a finite-dimensional approximation of the Koopman operator $\mathcal{K}$ from data by using a dictionary of observable functions that serves as a map from the state space to a feature space \cite{Tu2014}. Let the vector $\vec{\xi}\left(\vec{x}\right)$ be a dictionary of scalar observable functions that transform the state into new observables or features. In contrast to the DMD data matrices \eqref{eqn:dataMats}, EDMD forms the feature matrices \cite{Williams_2015}
\begin{equation}
\Xi =
\begin{bmatrix}
\bm{\xi}\left(\vec{x}_1\right)^T\\
\bm{\xi}\left(\vec{x}_2\right)^T\\
\vdots \\
\bm{\xi}\left(\vec{x}_N\right)^T\\
\end{bmatrix}
\qquad\text{ and }\qquad
\Xi' =
\begin{bmatrix}
\bm{\xi}\left(\vec{x}^{\prime}_1\right)^T \\
\bm{\xi}\left(\vec{x}^{\prime}_2\right)^T \\
\vdots \\
\bm{\xi}\left(\vec{x}^{\prime}_N\right)^T \\
\end{bmatrix},
\label{eqn:featureMats}
\end{equation}
and studies the operator $K_\text{EDMD}$ such that $\Xi' \approx \Xi K_\text{EDMD}$.  Having a sufficiently rich dictionary $\bm{\xi}$ of observable functions is important, but the memory requirements greatly increase as additional functions are added.

\subsection{Kernel Dynamic Mode Decomposition}
To make EDMD accessible to high dimensional systems like those found in fluid mechanics, Kernel Dynamic Mode Decomposition (KDMD) employs the kernel trick from machine learning to avoid having to actually form the feature matrices \cite{Williams_2015}. To illustrate the KDMD process, consider snapshot pairs $(\vec{x}_j,\vec{x}'_j)$ for $j= 1, \dots, N$. KDMD forms auxiliary matrices $\hat{G}$ and $\hat{A}$ by evaluating a kernel function $f(\cdot , \cdot)$ using various pairings of data snapshots such that \cite{Williams_2015}
\begin{equation}
\hat{G}_{ij} = f(\vec{x}_i,\vec{x}_j), \;\;\;\; \hat{A}_{ij} = f(\vec{x}'_{i},\vec{x}_j), \;\;\;\; i,j \in 1,\cdots , N. \label{eq:G_A}
\end{equation}
The entries of the $\hat{G}$ and $\hat{A}$ matrices are inner products of the feature vector evaluated on snapshots supplied to the kernel function. The feature vector is never explicitly formed, yet results of the associated inner products populate the $\hat{G}$ and $\hat{A}$ matrices. A common kernel is the polynomial kernel \cite{Williams_2015}
\begin{equation}
f(\vec{x}_i,\vec{x}_j) = \left( 1 + \frac{\vec{x}_j^T\vec{x}_i}{d^2}\right)^\alpha ,
\label{eqn:polyKer}
\end{equation}
where $\alpha$ is the order of the polynomial kernel and $d$ is a scaling factor that is dependent on relevant length scales in the problem under study. The polynomial kernel provides a method to evaluate inner products for vectors represented in a feature basis consisting of polynomial terms up to (and including) order $\alpha$. Although polynomial kernel framework of \eqref{eqn:polyKer} is capable to higher-order polynomial kernels, this paper sets $\alpha=1$ and $d=1$. Other polynomial choices and kernel forms are possible but are not considered in this paper.

An eigendecomposition of the auxiliary matrix $\hat{G} = V\Sigma^2 V^T$ provides $V$ and $\Sigma$. A key observation in the KDMD method is that the non-zero eigenvalues of the approximate Koopman operator $K_\text{EDMD}$ are also eigenvalues of the matrix \cite{Williams_2015}
\begin{equation}
\hat{K}= (\Sigma^{\dagger} V^T) \hat{A} (V \Sigma^{\dagger}).
\end{equation}
Further, the eigenvectors of $\hat{K}$ are projected versions of the eigenvectors of $K_\text{EDMD}$ \cite{Williams_2015}. Let $\Theta$ and $\Lambda$ be matrices derived from the eigendecomposition $\hat{K} = \Theta \Lambda \Theta^T$. The matrix of Koopman eigenfunction values $\Phi_x$ is \cite{Williams_2015}
\begin{equation}
\Phi_x = V \Sigma \Theta,
\end{equation}
and the Koopman modes for the state observables $X$ derive from the best-fit
\begin{equation}
Q_x = \Phi_x ^{\dagger}X = \Theta^{-1} \Sigma^{\dagger} V^T X,
\end{equation}
where the Koopman modes lie in the rows of $Q_x$. For the data set $(X,X')$, the matrices $\Lambda$, $\Phi_x$, and $Q_x$ contain approximations to the Koopman eigenvalues, evaluated Koopman eigenfunctions, and Koopman modes, respectively. Surana and Banaszuk \cite{Surana_2016} refer to the collection of matrices $\left(\Lambda, \Phi_x, Q_x\right)$ as the Koopman tuple, which can be used to form a reconstruction of the state.  To produce a Koopman tuple truncated to rank $r$, truncated matrices $V_r$ and $\Sigma_r$ should be formed from columns and diagonal entries of the $V$ and $\Sigma$ matrices, respectively, that correspond to the $r$ largest values on the diagonal of $\Sigma$. 

\subsection{Koopman Observer Form} \label{subsec:KOF}

Koopman Observer Form is an arrangement of a system's dynamics into a linear representation that is useful for implementation of a dynamic observer \cite{Surana_2016}. In KDMD, system measurements constitute the snapshots in data matrices, and the Koopman modes correspond to all system observables. However, for the purpose of subsequent estimation using an observer or estimate, KOF distinguishes between state and output observables. Three types of observables are considered in this work. Flow-field components are considered state observables. State observables describe the state of the flow field and come from fluid experiments or simulations of the system. These observables are only available to the user during model construction. Functions of the state observables are dictionary observables. Dictionary observables derive from the EDMD algorithm, which considers transformations of the state vector. These observables are important for approximation of the Koopman operator. However, KDMD does not calculate the dictionary observables; it avoids explicitly forming the vector of dictionary observables through use of the kernel trick from machine learning. Lastly, output observables are quantities directly measured during the process of flow estimation. Output observables are available with state observables during model construction, but they are the only quantities available during estimation. For sensor placement, measurables from candidate sensors are treated as system outputs.

A linear model for the time-evolution of a state observables vector $\vec{x}_k$ can be obtained from \eqref{eqn:KMD}, by letting $g(\vec{x}_k) = \vec{x}_k$. In order to use only real-valued quantities in a dynamic model of the state observables, Surana and Banaszuk \cite{Surana_2016} carefully separate real and imaginary components. In place of the matrix of Koopman eigenvalues $\Lambda$, they define a dynamics matrix $F$ that is block-diagonal with diagonal entry $F_{j,j} = \lambda_j$ if $\lambda_j$ is real, and
\begin{equation}
\begin{bmatrix}
F_{j,j} & F_{j,j+1}\\
F_{j+1,j} & F_{j+1,j+1}
\end{bmatrix}
=
\begin{bmatrix}
\text{Re}(\lambda_j) & \text{Im}(\lambda_j)\\
-\text{Im}(\lambda_j) & \text{Re}(\lambda_j)
\end{bmatrix},
\end{equation}
if $\lambda_j$ is complex. Let $\vec{z}_k$ be a modal state vector with only real entries that evolve according to $\vec{z}_k = F \vec{z}_{k-1}$. The matrix of state Koopman modes $Q_x$ links the state vector to the modal state vector, but a new matrix $H_x$ must be formed to handle the complex conjugate mode pairs of $Q_x$ and to adjust their dimensional orientation. Let $H_{x,j}$ and $(Q^T_x)_j$ denote the $j$th columns of matrices $H_x$ and $Q^T_x$, respectively. To accomplish this adjustment, set
\begin{equation}
H_{x,j} = (Q_x^T)_j,
\label{eqn:Hx_real}
\end{equation}
if the corresponding $\lambda_j$ is real, and
\begin{equation}
H_{x,j} = 2\text{Re}\left((Q_x^T)_j\right) \;\;\;\;\; H_{x,j+1} = -2\text{Im}\left((Q_x^T)_j\right),
\label{eqn:Hx_complex}
\end{equation}
if the corresponding $\lambda_i$ and $\lambda_{i+1}$ form a complex conjugate pair.

Consider a vector of output observables $\vec{y}(\vec{x})$ and arrange measurements of these observables in a data matrix $Y$ that is of the same form as the data matrices in \eqref{eqn:dataMats}. Output Koopman modes are regressions of the output measurements onto the Koopman eigenfunctions that are contained in the rows of the matrix
\begin{equation}
Q_y = \Phi_x ^{\dagger}Y^T = \Theta^{-1} \Sigma^{-1} V^T Y^T.
\end{equation}
An output matrix $H_y$ can be formed from $Q_y$ using the same procedure that formed $H_x$ from $Q_x$ in \eqref{eqn:Hx_real} and \eqref{eqn:Hx_complex}. In Koopman Observer Form, the system dynamics are \cite{Surana_2016}
\begin{subequations}
\begin{align}
\vec{z}_k &= F \vec{z}_{k-1} \label{eq:KOF1}\\
\vec{x}_k &= H_x \vec{z}_k \label{eq:KOF2}\\
\vec{y}_k &= H_y \vec{z}_k.\label{eq:KOF3},
\end{align}
\label{eq:KOF}
\end{subequations}
which represent a linear, real-valued model of the system that was built by applying KDMD to pairs of data snapshots.

\subsection{Discrete-Time Kalman Filter and Steady-State Kalman Filter}
The Koopman Observer Form \eqref{eq:KOF} provides an approximate dynamic model that is built from data. Sources of approximation include the use of a finite number of Koopman eigenfunctions (i.e., the finite size of the vector $\vec{z}_k$) and the choice of the dictionary of observables in EDMD or kernel in KDMD. A model built from noisy data also contains error due to external disturbances. For the purpose of state estimation, external disturbances may be included in model \eqref{eq:KOF} to yield the linear, stochastic representation
\begin{subequations}
\begin{align}
\vec{z}_k &= F \vec{z}_{k-1} + \vec{w}_{k-1}\\
\vec{y}_k &= H_y \vec{z}_k + \vec{v}_k,
\end{align}
\label{eqn:discrdyn}
\end{subequations}
where $\vec{w}_{k-1}$ is a process noise realization at time step $(k-1)$, and $\vec{v}_k$ is a measurement noise realization at time step $k$. By assumption, these noise processes are uncorrelated and Gaussian distributed such that $\vec{w}_{k-1} \sim {\cal N}\left(\vec{0},Q\right)$ and $\vec{v}_k \sim {\cal N}\left(\vec{0},R\right)$, where $Q$ and $R$ are covariance matrices. The process noise covariance $Q$ and measurement noise covariance $R$ can be computed from data using \cite{Gomez_2019}
\begin{subequations}
\begin{equation}
Q = \text{cov}(H_x^\dagger \vec{x}_{k+1} - F H_x^\dagger \vec{x}_k)
\end{equation}
and
\begin{equation}
R = \text{cov}(\vec{y}_k - H_y H_x^\dagger \vec{x}_k),
\end{equation}
\end{subequations}
respectively. 

The discrete-time Kalman filter can estimate the state of the stochastic dynamical system \eqref{eqn:discrdyn} from output measurements, provided that the pair $(F,H_y)$ is observable, which means that the time evolution of measurements provides sufficient information from which to infer the initial state of the system. Moreover, the discrete-time Kalman filter is the optimal filter for a discrete-time linear dynamical system with uncorrelated Gaussian process and measurement noise. The estimate $\hat{z}_k$ produced by the discrete-time Kalman filter can be converted to an estimate of $\hat{x}_k$ via \ref{eq:KOF2}.  Algorithm 1 provides the discrete-time Kalman filter \cite{Simon_Book_CH5}.  The notation $\hat{(\cdot)}$ denotes an estimated quantity. The superscript $(\cdot)^f$ denotes a forecasted quantity that does not include assimilation of data at the current time step, and the superscript $(\cdot)^a$ denotes that the quantity includes assimilation of data available at the current time step.

{\setstretch{1.0}
\begin{tcolorbox}[breakable, enhanced, title=Alg. 1: Discrete-Time Kalman Filter]
\noindent Inputs: Initial state estimate $\hat{z}_0^{a}$ and initial covariance $P_0^{a}$.
\begin{enumerate}
\begin{subequations}
\item[1.1)] Time update:
\begin{align}
\hat{\vec{z}}_k^{f} &= F\hat{\vec{z}}^a_{k-1}\\
P_k^{f} &= F P^a_{k-1} F^T + Q\label{eqn:covUpdate}
\end{align}
\item[1.2)] Kalman gain:
\begin{equation}
K_k = P_k^fH_y^T(H P_k^f H_y^T + R)^{-1}
\end{equation}
\item[1.3)] Measurement assimilation:
\begin{align}
\hat{\vec{z}}^a_k &= \hat{\vec{z}}_k^f + K_k(\vec{y}_k-H_y\hat{\vec{z}}_k^f)\\
P^a_k &= (I - K_kH_y)P_k^f \label{eqn:cov_apriori}
\end{align}
\end{subequations}
\end{enumerate}
\noindent Output: Estimate $\hat{\vec{z}}_k^a$
\end{tcolorbox}
}

If the discrete-time dynamical system \eqref{eqn:discrdyn} yields a stable Kalman filter, in the limit as $k \rightarrow \infty$, the Kalman gain matrix $K_k$ and the state covariance matrices $P_k^f$ and $P_k^a$ converge to steady-state matrices $K_\infty$, $P_\infty^f$, and $P_\infty^a$, respectively. Historically, the convergence of the discrete-time Kalman filter to a steady-state filter has been exploited to create steady-state filters that are capable of very rapid implementation in hardware \cite{Simon_Book_CH5}. A steady-state Kalman filter uses the steady-state Kalman gain matrix $K_\infty$ in place of $K_k$ in the discrete-time Kalman filter. This choice of gain is typically suboptimal during an initial transient period, but the difference in performance is often minor and is justified by the computational savings \cite{Simon_Book_CH5}.

The steady-state forecast covariance $P_\infty^f$ can be obtained by combining \eqref{eqn:covUpdate} and \eqref{eqn:cov_apriori} to yield the Discrete Algebraic Riccati Equation (DARE)
\begin{equation}
P_\infty^f = F P_\infty^f F^T - F P_\infty^f H_y^T \left(H_y P_\infty^f H_y^T + R\right)^{-1}H_yP_\infty^f F^T + Q.
\label{eqn:covRiccati}
\end{equation}

Although $P_k^f$ in Alg. 1 generally varies in time and depends on the initial condition and initial uncertainty, $P_\infty^f$ does not. The state-steady covariance is only a function of the noise covariances $Q$ and $R$, the dynamics matrix $F$, and the measurement matrix $H_y$. That is, $P_\infty^f$ is intrinsic to the system under study and independent of the initial conditions and initial uncertainty. These observations also hold for the {\it a posteriori} steady-state covariance $P_\infty^a$, which can be found from $P_\infty^f$, $K_\infty$, and \eqref{eqn:cov_apriori}. The Riccati equation \eqref{eqn:covRiccati} can also be written in terms of an information matrix. Using \eqref{eqn:covUpdate} at steady-state and the definition of the {\it a posteriori} information matrix as ${\cal I}^a_\infty = \left(P_\infty^a\right)^{-1}$, the Riccati equation \eqref{eqn:covRiccati} leads to \cite{Sagnol_2015}
\begin{equation}
{\cal I}^a_\infty = \left(F \left({\cal I}^a_\infty\right)^{-1} F^T + Q\right)^{-1} + H_y^T R^{-1}H_y.
\label{eqn:infoRiccati}
\end{equation}
Equation \eqref{eqn:infoRiccati} is an equivalent {\it a posteriori} form of \eqref{eqn:covRiccati}. Since the steady-state matrices are intrinsic to the system, they can be used for sensor design that is applicable to many different initial conditions. Section \ref{sec:Sensor} considers how the steady-state information matrix $I_\infty^a$ changes due to changes in the sensors included in the system's output.

%%%%%%%%%%%%%%%%%%%%%%%%%%%%%%%%%%%%%%%%%%%%%%%%%%%%%%%%%%%%%%%%%%%%%%%%%%%%%%%%%%%%%%%%%%%%%%%%%%%%%%%%%%%%%%%%%%%%

\section{Sensor Array Design for the Steady-State Kalman Filter} \label{sec:Sensor}
This section designs sensor arrays for the purpose of estimating unsteady flow fields using the data-driven modeling and estimation framework from Sec. \ref{sec:Background}. To accomplish array design, a resource allocation method from \cite{Sagnol_2015} provides optimal resource weights to maximize an information measure for a steady-state Kalman filter. The resource allocation weights sensors in an array of candidate sensors, but the solution is not sparse. To address this issue, this section proposes two methods to sequentially select sensors for a steady-state Kalman filter based on optimal resource allocation weights.

\subsection{Resource Allocation for a Steady-State Kalman Filter}
\label{sec:resAlloc}
Sagnol and Harman \cite{Sagnol_2015} pose a resource allocation problem to maximize a scalar function $\Phi({\cal I}^a_\infty)$ of the information matrix for a steady-state Kalman filter by optimally distributing sensing resources. Let $\vec{\omega} = [\omega_1,\omega_2, \dots,\omega_s]$ be a vector of resource allocation weights with $\omega_j \geq 0$ for sensor numbers $j=1,\dots,s$. To model the allocation of resources to a sensor, Sagnol and Harman assume that the measurement noise variance for a sensor depends on its allocation weight such that the measurement noise covariance matrix $R$ is diagonal with entries \cite{Sagnol_2015}
\begin{equation}
\sigma_{j}^2 = \frac{1}{\mu_j\left(\omega_j\right)},
\end{equation}
where $\mu_j(\omega_j) : \mathbb{R}_{\geq 0} \rightarrow \mathbb{R}_{\geq 0}$ is a continuous, nondecreasing, concave map with $\mu_j(0) = 0$. In the current paper, we select $\bm{\mu}(\vec{\omega}) = \vec{\omega}$ and set $\Phi({\cal I}^a_\infty) = \sqrt[n]{\text{det } {\cal I}^a_\infty}$, which is a common performance measure in D-optimal design \cite{Pukelsheim_2006}.

The weights in $\vec{\omega}$ represent how resources are allocated to the sensors to maximize the information measure $\Phi(X)$. A higher weight indicates that the corresponding sensor can provide more information towards maximizing $\Phi(X)$. As a resource approaches zero, the variance goes to infinity, indicating that the sensor does not provide useful information. To model resource constraints we choose, the set of admissible allocations to be the $s$-dimensional simplex
\begin{equation}
\Delta_s = \left\{ \vec{\omega} \in \mathbb{R}^s : \omega_j \geq 0 \text{ for } j=1,\dots,s \, \text{ and } \, \sum_{j=1}^s \omega_j = 1 \right\},
\end{equation}
thereby ensuring that the sensors must share the total of the resource weights.

For square matrices $A$ and $B$, let $A \succeq B$ denote that $A - B$ is a positive, semi-definite matrix. For a system with process noise covariance matrix $Q$, let $Q = L L^T$ be the Cholesky decomposition. Sagnol and Harman \cite{Sagnol_2015} use Linear Matrix Inequalities (LMIs) to formulate the SDP
\begin{align}
\qquad&\qquad\quad\max_{\vec{\omega} \in \mathbbm{R}^s \atop X,U \in \mathbbm{S}_n} \Phi(X) \nonumber \\
\text{s.t.} \; \; \; &
\begin{bmatrix}
X - F^TUF & F^TUL \\
L^TUF & I - L^TUL
\end{bmatrix}
\succeq 0 \nonumber \\
& X = U + H_y^T \text{diag}\left(\bm{\mu}(\vec{\omega})\right) H_y\succeq 0\label{eq:sagnol} \\
& \vec{\omega} \in \Delta_s . \nonumber
\end{align}

Sagnol and Harman \cite{Sagnol_2015} show that solving the SDP \eqref{eq:sagnol} admits a unique solution $\left(\vec{\omega}^*,X^*, U^*\right)$ that also solves the {\it a posteriori} steady-state information equation \eqref{eqn:infoRiccati} with ${\cal I}_\infty^a = X^*$ under the resource allocation $\vec{\omega}^*$. This solution process also provides the {\it a priori} steady-state information matrix ${\cal I}_\infty^f = U^*$. This contribution is noteworthy because the SDP \eqref{eq:sagnol} can be solved efficiently using standard software packages that implement interior-point methods. Section \ref{sec:Results} solves this SDP using the convex-optimization package CVX \cite{CVX1,CVX2} in MATLAB.

In the work of Sagnol and Harman \cite{Sagnol_2015}, the measurement noise covariance, prior to including resource weights, is assumed to be the identity matrix, such that the measurement noise in each sensor channel has unit variance and is uncorrelated with the noise in other channels. To account for sensor-to-sensor variations in measurement noise variance, we modify the second LMI in \eqref{eq:sagnol} to be
\begin{equation}
X = U + H_y^T \text{diag}\left(\bm{\mu}(\vec{\omega})\right) R_d^{-1} H_y\succeq 0, \label{eq:sagnol_mod}
\end{equation}
where $R_d$ is a diagonal matrix that contains the diagonal entries in $R$. Section \ref{sec:Results} builds $R$ from data, and small, nonzero values can appear in off-diagonal elements. Extracting the diagonal matrix $R_d$ from $R$ removes these spurious correlations, and the $R_d^{-1}$ term in \ref{eq:sagnol_mod} applies the appropriate scaling if the sensor noise is not unit variance in each output channel.

\subsection{Orthogonal Selection} \label{sec:orth_sel}
The resource allocation problem for a given array can be adapted to a sensor selection problem by including all candidate sensors such that the number of sensors for resource allocation becomes $s=n_c$. The assigned resource weights can then guide sensor selection.  An output that receives a large weight from the resource allocation problem presented in Section \ref{sec:resAlloc} provides important contributions in maximizing $\Phi({\cal I}_\infty^a)$. To develop a sequential method for sensor array design, consider the selection of the candidate sensor that achieves the largest weight in the resource allocation problem by adding the sensor to a set of selected sensors. The selected sensor could be removed from consideration, and the resource allocation problem could be re-calculated to continue the selection process. However, simply removing the selected sensor from consideration (e.g., by imposing a constraint that the corresponding weight be zero while solving the SDP in \eqref{eq:sagnol}) is insufficient for selection of diverse sensors. For example, consider a fine spatial discretization of candidate pressure sensors on the surface of an airfoil. Very closely located candidate sensors would provide very similar pressure readings. If one of the sensors is selected and removed in the first step of the process, a neighboring sensor that provides nearly identical pressure data would be selected on the second step of the process.

To account for the contributions of sensors already in the selected set, let $\vec{h}_\ast^T$ be the row of $H_y$ that is associated with the most recently selected sensor. Then, the direction $\left(\vec{h}_\ast^T / \lVert\vec{h}_\ast \rVert\right)$ is associated with the selected sensor in the row space of $H_y$. Other sensors may have measurement row vectors that are aligned with this direction or provide new directions within the row space.

The measurement matrix $H_y$ is altered after selection of a sensor to subtract off the portion of each row that is aligned with the $\left(\vec{h}_\ast^T / \lVert\vec{h}_\ast \rVert\right)$ direction. Let $H_y^{(j)}$ be the matrix $H_y$ on iteration $j$ for $j=1,\dots,n_c$. The update to the $H_y^{(j)}$ after the selection of a row is
\begin{equation}
H_y^{(j+1)} = H_y^{(j)} - H_y^{(j)} \left(\frac{\vec{h}_\ast}{\lVert\vec{h}_\ast \rVert}\right) \left(\frac{\vec{h}_\ast^T}{\lVert\vec{h}_\ast \rVert}\right).
\end{equation}
Thereby, each subsequent selection step only considers the portions of the measurement row vectors that are orthogonal to the sensor contributions in the selected set. The row in $H_y$ corresponding to the previously selected sensor becomes a row of zeros and no longer influences the selection process, thereby avoiding the need to constrain the weight of the selected sensor to be zero in subsequent steps.

\subsection{Complementary Selection}
The orthogonal selection method described in Sec.\ \ref{sec:orth_sel} works well for choosing sensors that provide diverse information, but it limits the number of sensors in an array to the rank of $H_y$. Consider $H_y$ to be an $n_c \times r$ matrix, where $n_c$ is the number of candidate sensors, and $r$ is the size of the reduced-order model. Generally, when considering many candidate sensors and a reduced-order model, we have $r < n_c$. Since the contributions of a selected row of $H_y$ are subtracted from all remaining rows of $H_y$ at each iteration of orthogonal selection, the rank of $H_y$ decreases by one for each iteration. The rank of $H_y$ becomes zero after $r$ sensors are selected. It is therefore not possible to use orthogonal selection to build an array with more than $r$ sensors.

Another approach to sequential array construction is to retain the selected sensors in the resource allocation problem during subsequent selection steps. The weights of the selected sensors are set to zero through an additional constraint in the SDP that $\omega_j = 0$ for $j\in S$, where $S$ is the set of the selected indices. These selected sensors are retained in the SDP by including an additional $H_y^T P_S H_y$ term in the constraint on $X$ such that
\begin{equation}
X = U + H_y^T P_s R_d^{-1} H_y + H_y^T \text{diag}\left(\bm{\mu}(\vec{\omega})\right) R_d^{-1} H_y,
\end{equation}
where $P_s$ is a diagonal selection matrix with ones corresponding to selected sensors and zeros corresponding to remaining candidate sensors. This method promotes diversity of information during sensor selection but allows for more sensors than the rank of $H_y$, and the sensors are selected in a manner that complements the array's existing sensors without imposing an orthogonal approach to selection. A comparison of the arrays formed using the orthogonal and complementary selection methods proposed in the present section is examined in Section \ref{sec:Results}.

%%%%%%%%%%%%%%%%%%%%%%%%%%%%%%%%%%%%%%%%%%%%%%%%%%%%%%%%%%%%%%%%%%%%%%%%%%%%%%%%%%%%%%%%%%%%%%%%%%%%%%%%%%%%%%%%%%%%

\section{Sensor Placement for Several Simulated Operating Points}\label{sec:Results}
This section presents sensor array design using flow data generated via CFD simulations of flow past an airfoil at various angles of attack and Reynolds numbers. Pressure readings are recorded at candidate sensor locations along the surface of the airfoil and pressure sensor arrays are designed using orthogonal selection, complementary selection, and QR-pivoting, which is another sensor-placement method used for comparison. This section also introduces a method to choose an array design that maximizes the information measure across several operating conditions. The resulting arrays are compared based on filtering error on a task of flow estimation.

\subsection{Computational Fluid Dynamics Simulations}\label{subsec:CFD}
All CFD simulations were performed using the commercial flow solver COMSOL Multiphysics, version 6.0 \cite{COMSOL}. COMSOL solved for the flow of water past a stationary NACA0012 airfoil of chord length $10.16$ cm ($4.00$ in) at angles of attack $30^\circ$, $35^\circ$, and $40^\circ$ and Reynolds numbers $3\times 10^3$, $4\times 10^3$, and $5\times 10^3$ for a total of nine operating points enumerated in Table \ref{tbl:op_points}.

\begin{table}[ht]
\caption{\label{tbl:op_points} Operating Points}
\small
\centering
\begin{tabular}{l|l l l }
Operating Pt. & 1 & 2 & 3 \\
\hline
$(\text{Re},\text{angle})$ & $(3 \times 10^3, 30^\circ)$ & $(4 \times 10^3, 30^\circ)$ & $(5 \times 10^3, 30^\circ)$ \\ 
\noalign{\smallskip}
\noalign{\smallskip}
\noalign{\smallskip}
& 4 & 5 & 6 \\
\cline{2-4}
& $(3 \times 10^3, 35^\circ)$ & $(4 \times 10^3, 35^\circ)$ & $(5 \times 10^3, 35^\circ)$ \\
\noalign{\smallskip}
\noalign{\smallskip}
\noalign{\smallskip}
& 7 & 8 & 9 \\
\cline{2-4}
& $(3 \times 10^3, 40^\circ)$ & $(4 \times 10^3, 40^\circ)$ & $(5 \times 10^3, 40^\circ)$ \\
\noalign{\smallskip}
\noalign{\smallskip}
\noalign{\smallskip}
\end{tabular}
\end{table}

\begin{figure}[t]
\centering
\subfigure[Geometry \label{fig:geometry}]
{
\includegraphics[width=.54\linewidth, clip=true, trim=0in 0in 0in 0in]{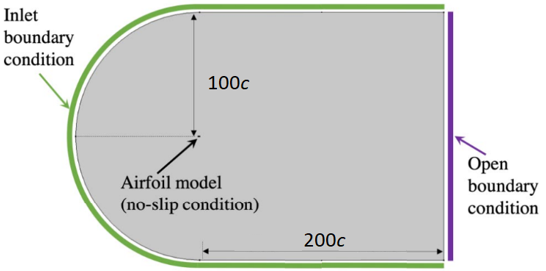}
}
\subfigure[Computational mesh \label{fig:mesh}]
{
\includegraphics[width=.42\linewidth, clip=true, trim=0in 0in 0in 0in]{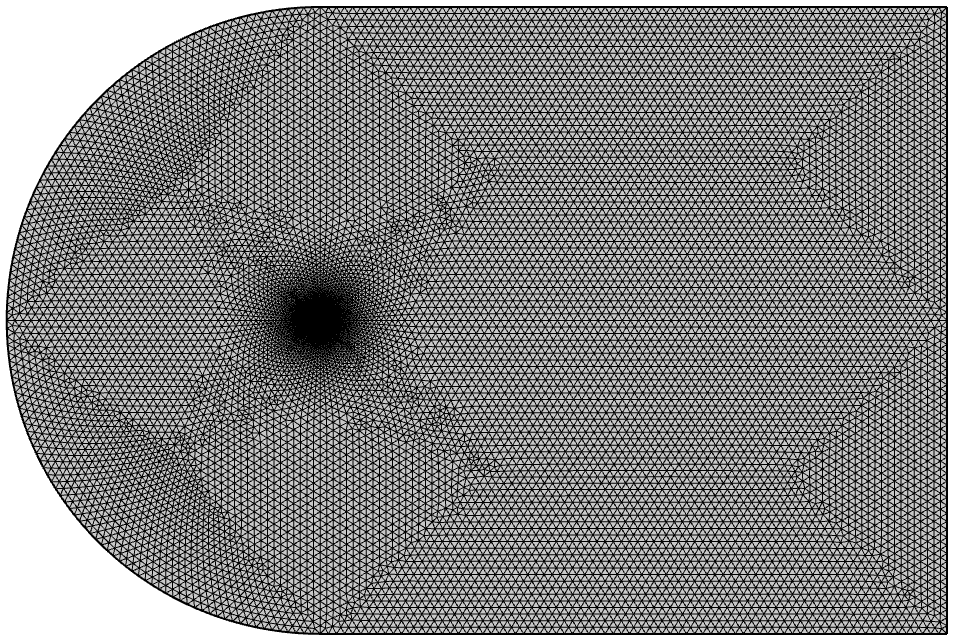}
}
\caption{Setup for COMSOL simulations}
\label{fig:COMSOL_setup}
\end{figure}

Solutions were obtained using COMSOL's laminar flow solver, which implements direct numerical simulation of the Navier-Stokes equations. After simulation, pressure signals collected from candidate sensor locations were corrupted by Gaussian, white noise with a standard deviation of $5\%$ of the mean signal value across all sensors. The geometry and boundary conditions of the COMSOL simulations are shown in Fig. \ref{fig:geometry}. The angle of attack of the airfoil was changed by altering the angle of the flow on the inlet boundary. Figure \ref{fig:mesh} contains the computational mesh used for the simulations, which consists of numerous boundary layers near the surface of the airfoil and free-triangular elements filling in the rest of the domain. Mesh refinement was performed until the unsteady lift force on the airfoil converged to a repeatable signal, independent of mesh density. 

In order to remove initial transient behavior, all simulations were run for 1000 seconds and only data from 825s to 975s were retained for analysis. The first 33\% of the retained data were only used in the flow feature analysis of Sec.\ \ref{sec:flow_features}. The next 47\% of retained data were used to train a KDMD model for the corresponding operating point, and the last 20\% of the retained data were used for filtering tests in Sec.\ \ref{subsec:compare}.

To determine the rank $r$ of the number of modes used in formation of dynamic models at each operating (i.e., the $F$, $H_x$, and $H_y$ matrices in \eqref{eq:KOF} at each operating point), analysis of the POD modal energy content was performed.   The selected rank was set to $r=6$ for models at all operating points in order to match the operating point which required $6$ POD modes to represent 99.88\% of the mean-subtracted flow's energy.  In subsequent design of sensor arrays, $n_s=6$ sensors were used for each array in an effort to match the rank of the model for the operating point with the most POD modes required for 99.88\% of the mean-subtracted flow's energy.

\subsection{QR-based Sensor Selection}

Manohar et al. \cite{QR_Manohar} proposed a data-driven method for sensor placement based on pivoted QR factorization of a basis matrix that is specific or tailored to the problem, such as a matrix of POD modes for a particular fluid flow. Since their method is entirely data-driven and has been widely adopted, we use it as a useful point of comparison for the methods proposed in this paper.

Consider the state data matrix $X$ from \eqref{eqn:dataMats}. Let $\Psi_r$ be a matrix containing POD modes of $X^T$, which are left-singular vectors resulting from the SVD of $X^T$ truncated to rank $r$ \cite{Lumley_1993}. Each state snapshot $\vec{x}_k$ can be reconstructed from the POD modes using a vector of appropriate weights, such that $\vec{x}_k = \Psi_r \vec{b}_k$. Manohar et al. \cite{QR_Manohar} solve the sensor placement problem in which candidate sensors are direct measurements of the state components by using the pivots from the QR decomposition of $\Psi_r^T$ to reveal the rows of $\Psi_r$ that are most important to approximating $\vec{b}_k$. The state components corresponding to these pivots are the sensors that contribute the most to the approximation of $\hat{\vec{b}}_k$ and its associated state reconstruction $\hat{\vec{x}}_k = \Psi_r \hat{\vec{b}}_k$. For cases where the desired number of sensors exceeds the rank $r$ of $\Psi_r$, which Manohar et al. \cite{QR_Manohar} refer to as the oversampled case, QR decomposition should be performed on $\Psi_r \Psi_r^T$. The current paper considers sensors that may not be state components, so it is necessary to append the candidate sensor measurements onto the state in the combined data matrix
\begin{equation}
\bar{X} =
\begin{bmatrix}
\vec{x}_1^T & \vec{y}_1^T \\
\vec{x}_2^T & \vec{y}_2^T \\
\vdots      & \vdots       \\
\vec{x}_{n_t}^T & \vec{y}_{n_t}^T \\
\end{bmatrix}
\end{equation}
The SVD of $\bar{X}^T$ provides a set of POD modes $\bar{\Psi}$ for the augmented system. Only the bottom left $n_c \times r$ corner of $\bar{\Psi}$ should be used in the QR decomposition, which corresponds to the portions of the $r$ most energetic POD modes that are associated with the candidate outputs $\vec{y}_k$. Let $\bar{\Psi}_{y}$ be the $n_c \times r$ rectangular matrix extracted from the lower left corner of $\bar{\Psi}$. Pivot selection via QR decomposition can be performed on $\bar{\Psi}_{y}^T$ for the case of $r$ sensors and on $\bar{\Psi}_{y}\bar{\Psi}_{y}^T$ for the over-sampled case. The members of $\vec{y}_k$ identified by the resulting pivot locations form the QR-based sensor arrays.

\subsection{Sensor Array Design across Operating Points} \label{subsec:array_results}
Orthogonal selection, complementary selection, and QR-pivoting sensor placement were each used to build sensor arrays from the simulated data described in Sec.\ \ref{subsec:CFD}. In addition, this subsection examines how to select a sensor array that performs well across a range of operating conditions.

One approach to array design for various conditions is to include the data for all operating points during sensor selection. To implement this approach, let $m$ be the number of different operating or design points.  Consider a virtual, composite system of several duplicate airfoils, with identical sensors, each operating at a different flow condition.   Let $\mathcal{I}_{\infty}^{a}(i,j)$ denote the steady-state, {\it a posteriori} information matrix associated with a Kalman filter performing flow estimation at operating point $i$ given sensor-array configuration $j$. The sensor array is assumed to know the correct flow model to employ at each evaluation point.  Since the measurements from different operating points are uncorrelated, the information matrix for this virtual, composite system is the block diagonal matrix
\begin{equation}
\mathcal{I}_{\infty}^{a}(\text{comp},j) = \text{diag}\left(\mathcal{I}_{\infty}^{a}(1,j), \; \mathcal{I}_{\infty}^{a}(2,j), \; \cdots , \; \mathcal{I}_{\infty}^{a}(m,j)\right),
\label{eq:block_diag}
\end{equation}
which is made up of the information matrices from the operating points.  For an information matrix $\mathcal{I}_{\infty}^{a}$ of size $d \times d$, selection of $\Phi(\mathcal{I}_{\infty}^{a}) = \sqrt[d] {\det {\cal I}_{\infty}^{a}}$ as the scalarization function in SDP \eqref{eq:sagnol} allows the information measure for the virtual, composite system to break apart. Note that the information matrix \eqref{eq:block_diag} for the composite system is size $nm \times nm$. Applying the measure $\Phi$ and invoking the property that the determinant of a block-diagonal matrix equals the product of the determinants of the blocks yields,
\begin{align}
\Phi\left({\cal I}_{\infty}^{a}(\text{comp},j)\right) &= \sqrt[m \cdot n]{ \det {\cal I}_{\infty}^{a}(\text{comp},j)}\nonumber\\
&=\sqrt[m]{\sqrt[n] {\det {\cal I}_{\infty}^{a}(1,j)} \sqrt[n] {\det {\cal I}_{\infty}^{a}(2,j)} \cdots \sqrt[n] {\det {\cal I}_{\infty}^{a}(m,j)}} \nonumber\\
&= \sqrt[m]{\Phi\left({\cal I}_{\infty}^{a}(1,j)\right) \Phi\left({\cal I}_{\infty}^{a}(2,j)\right) \cdots \Phi\left({\cal I}_{\infty}^{a}(m,j)\right)} . \label{eqn:CIM}
\end{align}
Therefore, selecting the array that solves
\begin{equation}
\max_i \sqrt[m]{\prod_{j=1}^{m} \Phi(\mathcal{I}_{\infty}^{a}(i,j))},
\label{eq:max_info}
\end{equation}
maximizes the information of the composite system. Note that \eqref{eq:max_info} contains the geometric mean of the information measures from the subsystems of the composite system. Further, the scalarization function is the geometric mean of the eigenvalues of the information matrix for the subsystems and for the composite system. The resource allocation problem for the composite system is solved using block diagonal matrices for $H$, $F$, $Q$, and $R$ using the same process shown in the construction of \eqref{eq:block_diag}, and enforcing that the weight vector $\vec{w}$ contain replications of the same weights across all operating points. This construction of $\vec{w}$ ensures that a sensor receives the same weighting at each operating point in the composite system. Equation \eqref{eq:max_info} gives the maximum information across all operating points so this becomes the objective for the resource allocation problem on the composite system. Orthogonal and complementary selection each generate the same sensor array from the composite system.

\begin{figure}[t]
\centering
\includegraphics[width=.98\linewidth, clip=true, trim=0in 0in 0in 0in]{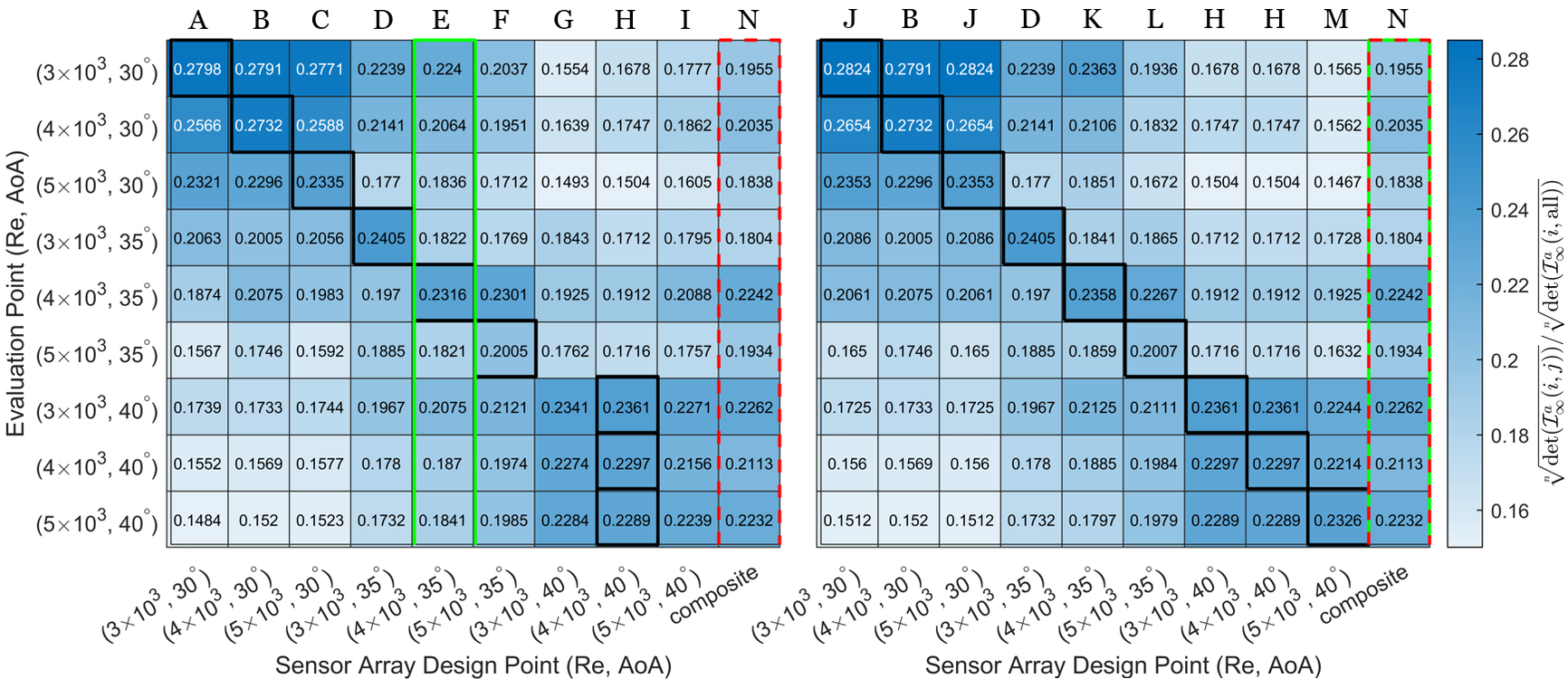}
\caption{Evaluation of array results based on a normalized measure of the steady-state information matrix (orthogonal selection left, complementary selection right)}
\label{fig:array_eval_information}
\end{figure}

Figure \ref{fig:array_eval_information} tabulates information measures for arrays at various operating points.  To examine information as a proxy for anticipated filtering performance of an array at various operating conditions, we distinguish between the design point, which is the operating condition for which and array was built, and the evaluation point, which is the operating condition at which the array is tested.  Figure \ref{fig:array_eval_information} analyzes arrays built using orthogonal selection for the left grid and complementary selection right grid. Each row corresponds to a different flow model generated using data from the listed evaluation point. Each column corresponds to a sensor array that was created for the listed design point. Each $(i,j)$ cell corresponds to a measure of the steady-state information matrix for the flow model of evaluation point $i$, using the array made for design point $j$. Since some operating conditions provide stronger output signals than others, normalization of the information measure is performed to allow comparison of information values from differing evaluation points (i.e. comparison of values in different rows). Each row $i$ is normalized by the maximum information measure possible for that operating condition $\Phi\left({\cal I}_{\infty}^{a}(i,\text{all})\right),$ which is the information obtainable if all candidate sensors are active. The rightmost column in each grid corresponds to the array designed using all the available flow models in a composite system. Diagonal entries report the information provided by array $j$ at the evaluation point for which it was designed. Off-diagonal entries assess the array's obtainable information at off-design conditions. Hence, each array design is evaluated at all possible evaluation points to consider performance across operating conditions.

\begin{figure}[t]
\centering
\includegraphics[width=.98\linewidth, clip=true, trim=0in 0in 0in 0in]{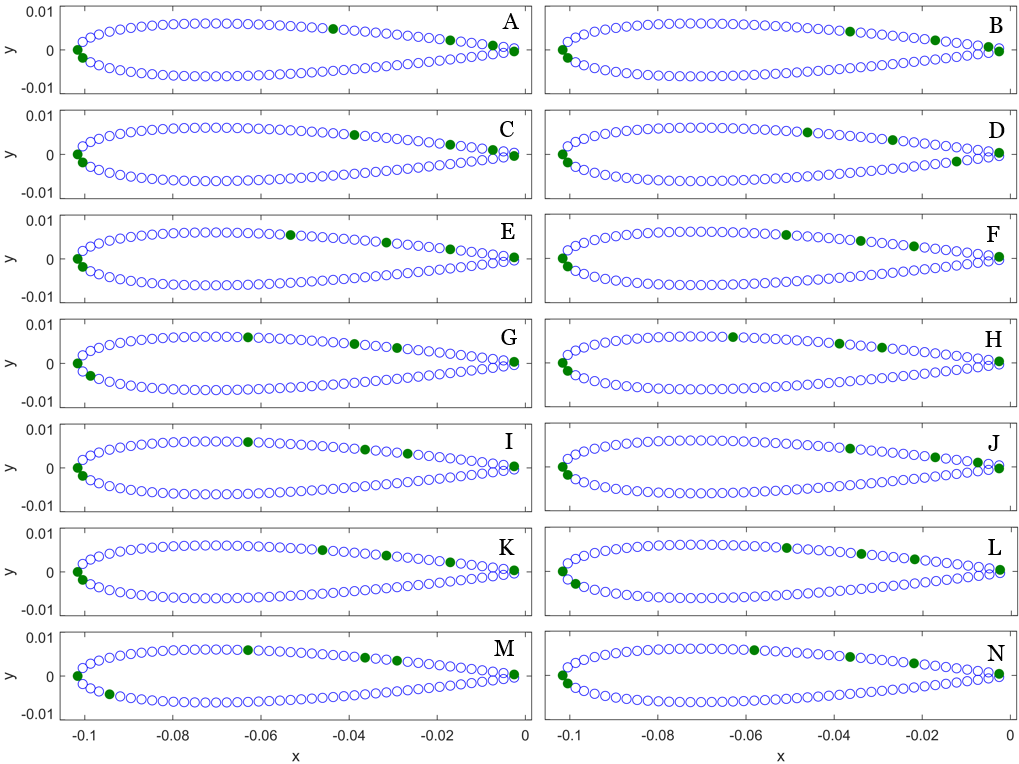}
\caption{Arrays constructed by orthogonal and complementary selection}
\label{fig:all_arrays}
\end{figure}

The array designs are identified by the letters at the top of each column in Fig. \ref{fig:array_eval_information}, with a letter assigned to each unique array design.  Sensor placement sometimes resulted in the construction of identical arrays from different design points and selection methods. Of the eighteen possible combinations of design points and selection methods, thirteen unique arrays were produced. The layouts of these arrays are shown in Fig. \ref{fig:all_arrays}. Candidate pressure sensor locations are blue circles, and selected sensors are filled-in green circles. Although some candidate locations near the thin trailing edge may not be manufacturable, they are included in this study for completeness. Such candidate locations can simply be discarded prior to running the selection algorithms if avoiding placement at these locations is desired.

In the complementary selection grid on the right of Fig. \ref{fig:array_eval_information}, notice that the values for design points $(\text{Re},\text{AoA}) = (4 \times 10^3, 30^\circ)$ and $(5 \times 10^3, 30^\circ)$ are identical, because identical arrays were formed for these design points. Similarly, information measures for design points $(\text{Re},\text{AoA}) = (3 \times 10^3, 40^\circ)$, and $(4 \times 10^3, 40^\circ)$ match due to matching arrays. For design points $(\text{Re},\text{AoA}) = (3 \times 10^3, 35^\circ)$, $(4 \times 10^3, 30^\circ)$, and $(4 \times 10^3, 40^\circ)$, the orthogonal and complementary selection methods constructed identical sensor arrays. Although identical arrays result for some design points, the order of sensor selection often differed during the design process.

The decomposition \eqref{eqn:CIM} shows that the information for a virtual, composite system for each array design can be found by taking the product of the entries in each column of Fig.\ \ref{fig:array_eval_information}. This information measure is proportional to the product of the eigenvalues of the composite system's steady-state information matrix for the corresponding array. To select an array design based on the anticipated information gain across all evaluation points, one would select the column that maximizes the product of its entries.  The column with the largest product is outlined in red.  The composite system provides the largest column product for both orthogonal and complementary selection. This result was expected since the column products are proportional to the information measure being maximized by objective function \eqref{eq:max_info}, which was used in the formation of the composite array. 

In Fig.\ \ref{fig:array_eval_information}, the cell with the maximum information measure in each row is outlined with a black border. Arrays designed at a specific angle of attack tend to have a high information measure when evaluated at the same angle regardless of flow speed. This finding suggests that sensor array design is more sensitive to changes in angle of attack than flow speed for these operating conditions. It is reasonable to expect that the maximum information value in each row should lie on the diagonal, since the design and evaluation points agree on the diagonal. For complementary selection, all of the largest values lie on the diagonal. For orthogonal selection, the maximum information measure for evaluation points $(\text{Re},\text{AoA}) = (3 \times 10^3, 40^\circ)$, and $(5 \times 10^3, 40^\circ)$ do not lie on the diagonal. In both of these instances, an array designed for the same angle of attack provided the maximum information measure for the evaluation point (i.e. for that row). Further, the discrepancy between the maximum information measure and the diagonal entry for that row is small in both cases. The cause is likely due to orthogonal selection's emphasis on the orthogonality of sensors rather than information maximization. Even for complementary selection, it may be possible to have an array from a different design point provide the most information at a specified evaluation point due to the sequential nature of the method, although it did not occur in this case.

\subsection{Filtering Results and Comparison of Methods} \label{subsec:compare}

\begin{figure}[t]
\centering
\subfigure[Designed and evaluated at $(4 \times 10^3$, $30^\circ)$ \label{fig:filter_point_2}]
{
\includegraphics[width=.48\linewidth, clip=true, trim=0in 0in 0in 0in]{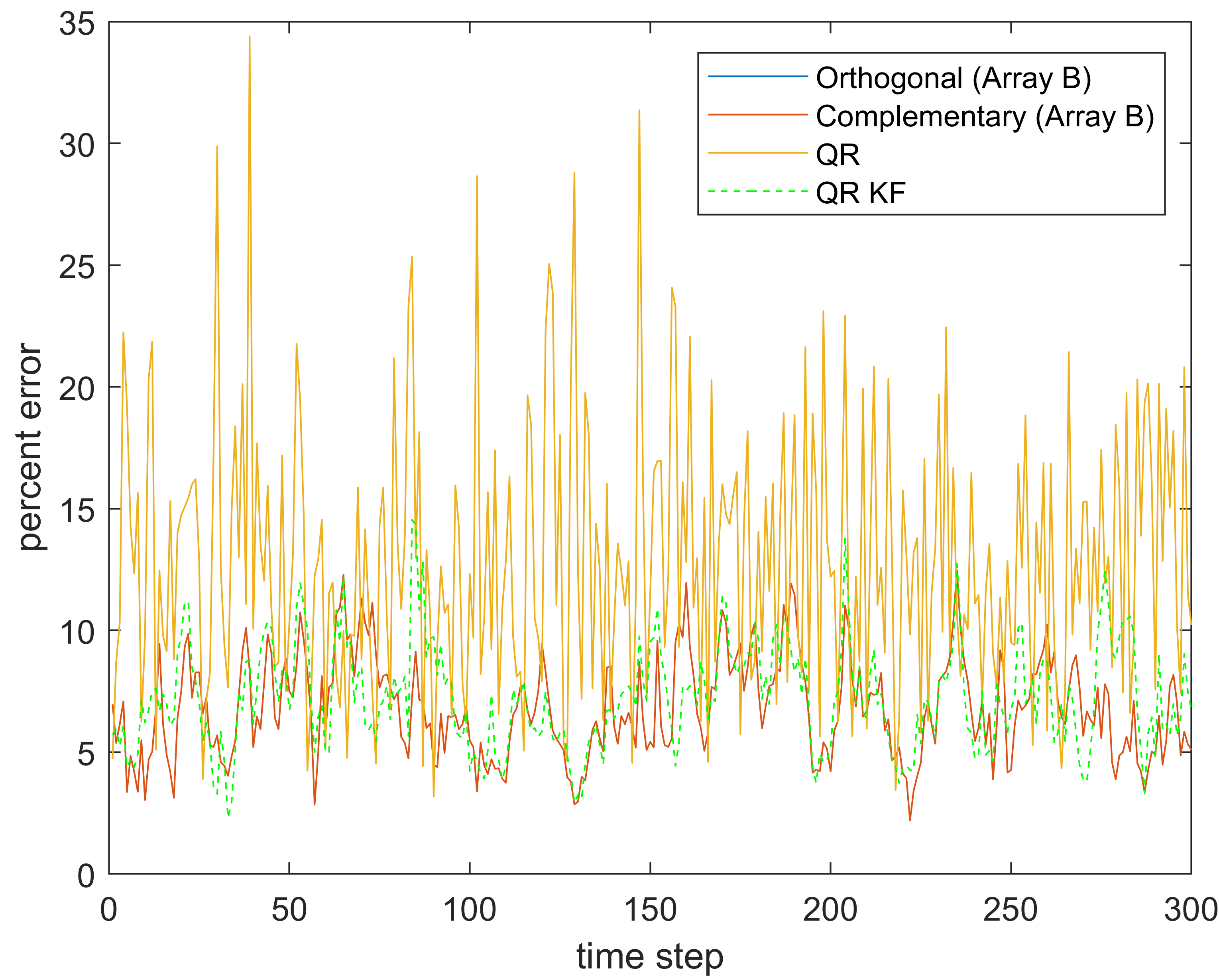}
}
\subfigure[Designed and evaluated at $(4 \times 10^3$, $35^\circ)$ \label{fig:filter_point_5}]
{
\includegraphics[width=.48\linewidth, clip=true, trim=0in 0in 0in 0in]{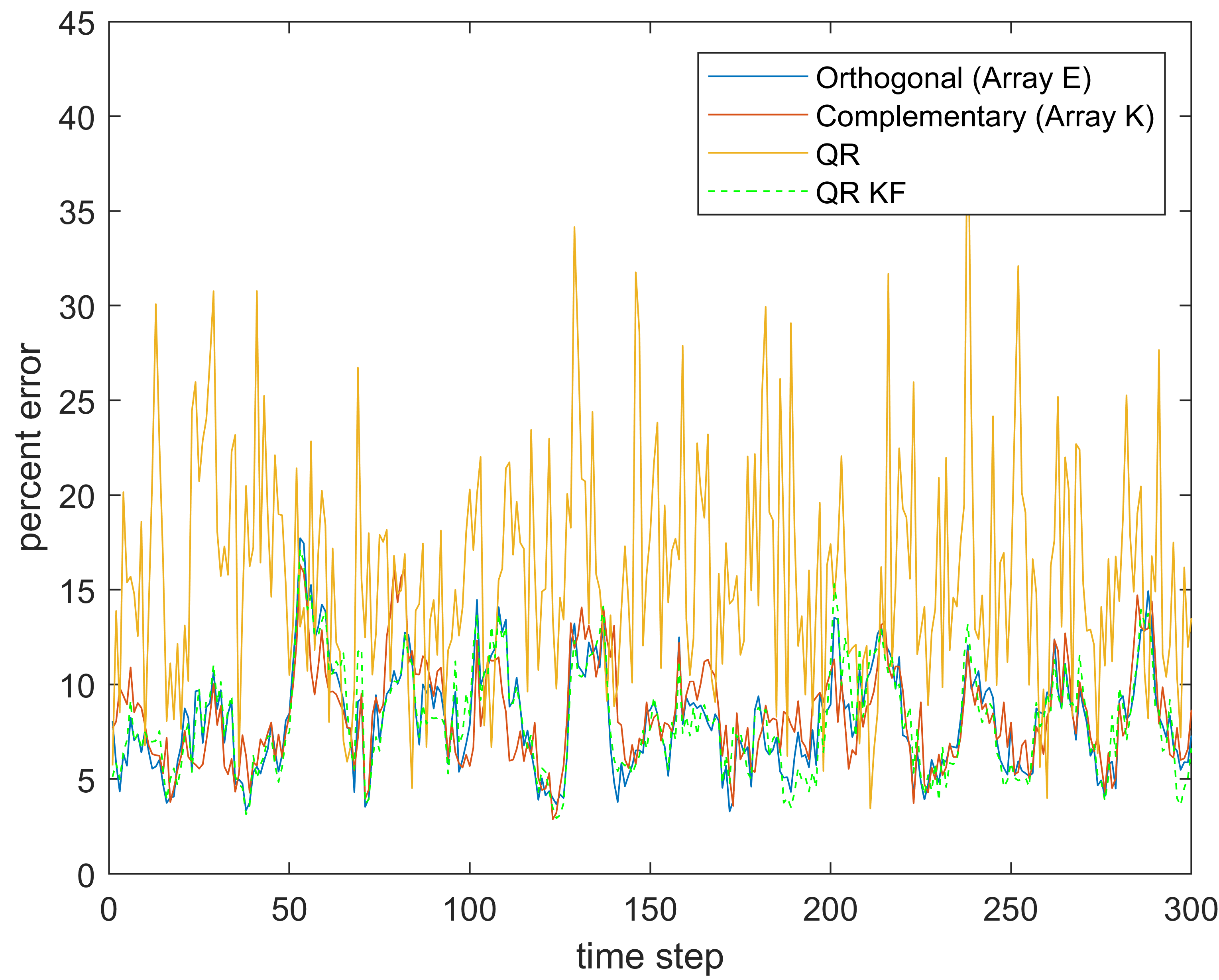}
}
\caption{Examples of time evolution of Kalman filtering results}
\label{fig:KF_individual}
\end{figure}

This section presents an analysis of filtering performance of the selected arrays. A sensor array has been designed using each of the three sensor placement methods at each of the nine design points. QR-pivoting sensor arrays are designed for instantaneous estimation of the flow field, while orthogonal and complementary selection arrays are designed for filtering estimation. Hence, instantaneous estimates of the flow are generated using the arrays built from QR-pivoting and Kalman filtering estimates of the flow are generated using the arrays built from the two information-based methods. Let $\vec{x}(t)$ be the measure flow field data at time $t$, $\hat{\vec{x}}(t)$ be the estimate if the flow field data at time $t$, and let $100 ||\vec{x}(t)-\hat{\vec{x}}(t)|| / ||\vec{x}(t)||$ be the \% error in the flow field estimate at time $t$. Figure \ref{fig:KF_individual} shows the time evolution of the estimation error for the three different sensor selection methods. We also included the error that results when the arrays constructed using QR-pivoting are permitted to use the data-driven flow models from our information-based sensor selection framework to generate Kalman filtering estimates of the flow field. The array design point and the evaluation point share the indicated operating point in Figs. \ref{fig:filter_point_2} and Fig. \ref{fig:filter_point_5}. Complementary selection and orthogonal selection both produced the same sensor array for design point $(4 \times 10^3, 30^\circ)$, resulting in overlapping error traces on Fig. \ref{fig:filter_point_2}. Since QR-pivoting array estimates are memoryless, the error associated with this method is of greater magnitude and choppier than the filtering estimates. Arrays built from orthogonal and complementary selection outperform the array built from QR-pivoting. However, this is primarily attributable to the difference in estimation methods. The QR arrays perform comparably well when they are permitted to generate Kalman filtering estimates using our KDMD models.

\begin{figure}[t]
\centering
\subfigure[Orthogonal selection array performance \label{fig:filter_Orth}]
{
\includegraphics[width=.48\linewidth, clip=true, trim=0in 0in 0in 0in]{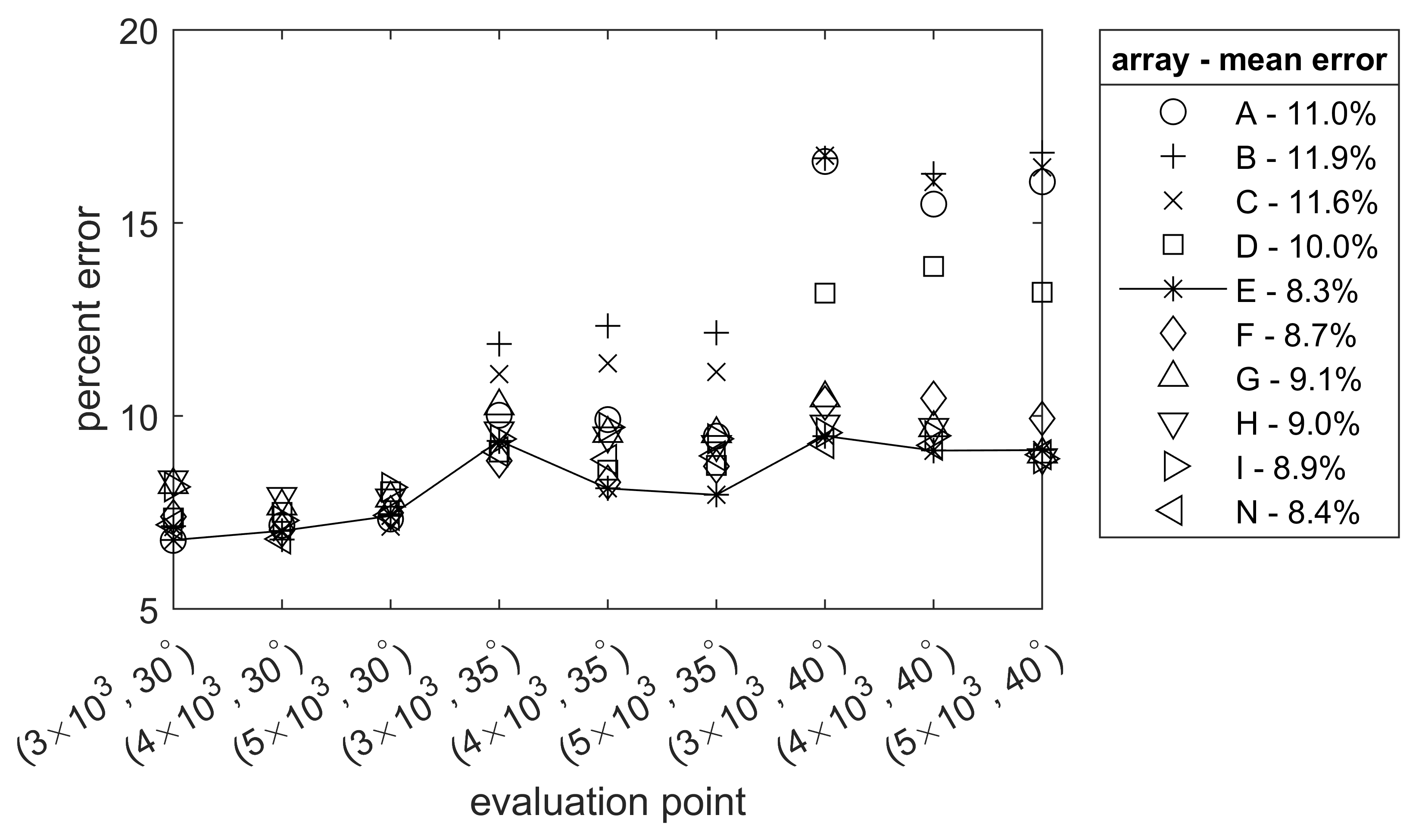}
}
\subfigure[Complementary selection array performance \label{fig:filter_Comp}]
{
\includegraphics[width=.48\linewidth, clip=true, trim=0in 0in 0in 0in]{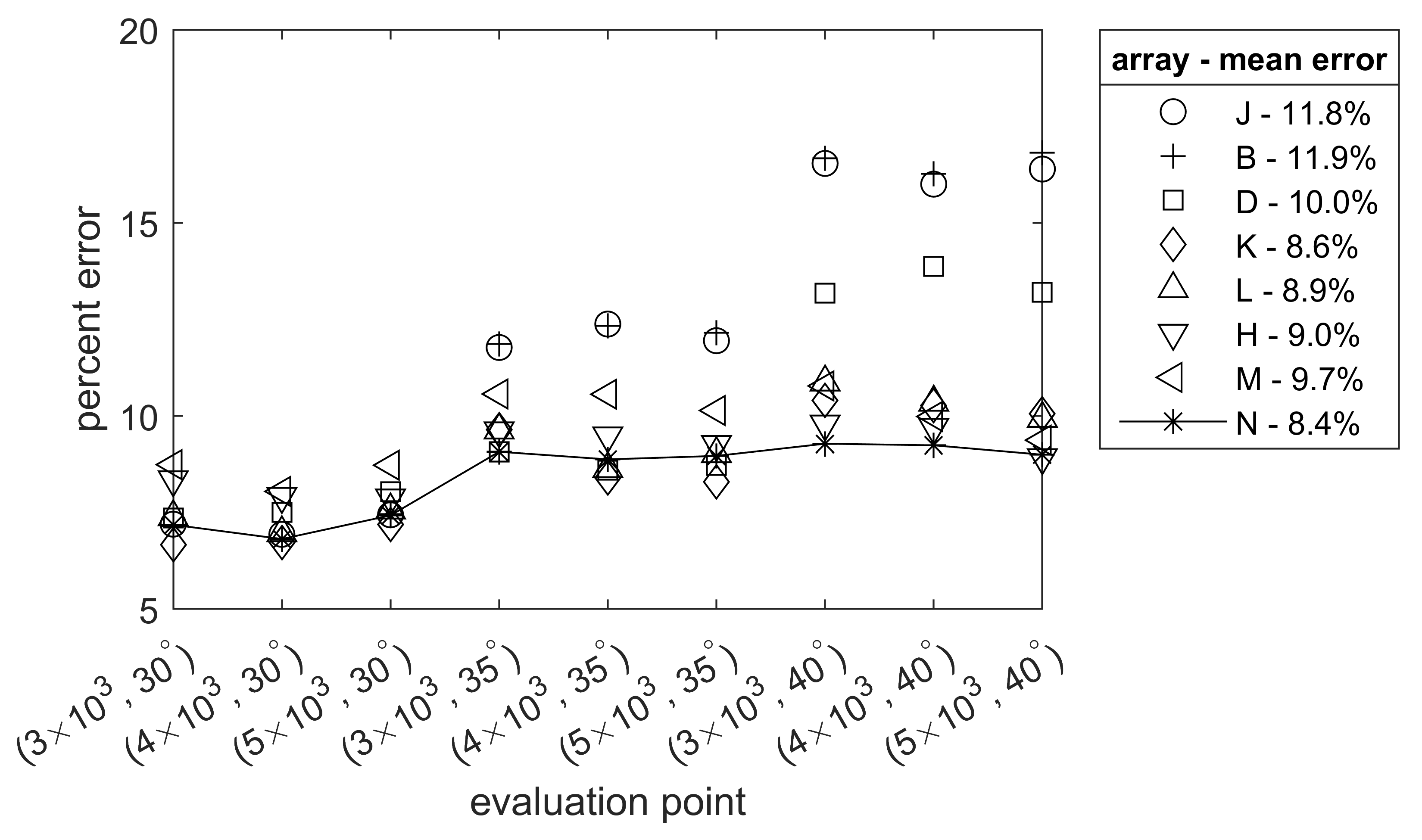}
}
\caption{Kalman filtering performance for each design-evaluation combination}
\label{fig:KF_results}
\end{figure}

To examine the relative performance of the sensor arrays in Fig.\ \ref{fig:all_arrays}, we compute the time-averaged filtering error at each evaluation point. Figure \ref{fig:KF_results} shows plots of the performance measures for arrays built using complementary selection and orthogonal selection. Sensor array information is positively correlated with sensor array filtering performance; however, this relationship is not linear, so evaluation points with slightly lower information can result in significantly lower filtering performance. This means that using the product of the columns of Fig.\ \ref{fig:array_eval_information} is not an ideal predictor of average filtering performance for an array. Instead, we propose ordering the arrays based on the minimum entry in each column of Fig.\ \ref{fig:array_eval_information}, where the predicted best filtering performance is provided by the array with the largest minimum information entry across all evaluation points. For each selection method in Fig.\ \ref{fig:array_eval_information}, the column with the largest minimum entry is outlined in green. The corresponding arrays in Fig.\  \ref{fig:KF_results} are shown by the marker $\ast$ and are connected with a line. Since these two arrays were expected to have the best average filtering performance, most of the points on the plot should lie above the lines for these arrays. The percent error for each array, averaged over all evaluation points, is provided in the legends of Fig.\ \ref{fig:KF_results}. The arrays with the lowest average filtering error agree with the arrays selected with green outlines in the information analysis of Fig.\ \ref{fig:array_eval_information}. For both selection methods, choosing the array that maximizes the minimum information measure resulted in the best average filtering performance across operating conditions. Orthogonal selection slightly outperforms complementary selection when looking at the mean error for the best performing arrays. Interestingly, the array built using all of the data in a composite system (i.e. array N) was not the best-filtering array. Instead, orthogonal selection was able to produce array E that gave the best filtering performance. Array E outperforming array N can be attributed to the fact that the sequential selection methods provided in this paper do not exhaustively consider all possible sensor combinations. Also note that the operating condition used to build array E was ($4 \times 10^3, 35^\circ$), which is the average flow speed and average angle of attack over all conditions. This finding suggests that obtaining flow data at an average operating condition may be sufficient for building a sensor array that can work well in a range of operating conditions, provided that there are no significant changes in flow behavior across conditions.

In general, orthogonal selection and complementary selection performed comparably well. There is an upper limit on the total number of sensors that can be selected using orthogonal selection, up to the rank $r$ of the $H$ matrix. In addition, the $H$ matrix can become ill-conditioned during orthogonal selection due to the adjustments made to the rows of the $H$ matrix, even before $r$ sensors have been selected. Due to these limitations of orthogonal selection and the comparable performance between the two methods, we recommend the use of complementary selection.

\section{Analysis of Sensor Locations and Nearby Flow Features} \label{sec:flow_features}
\begin{figure}[t]
\centering
\subfigure[Surface pressure and stagnation points \label{fig:physics_st_pt}]
{
\includegraphics[width=.48\linewidth, clip=true, trim=0in 0in 0in 0in]{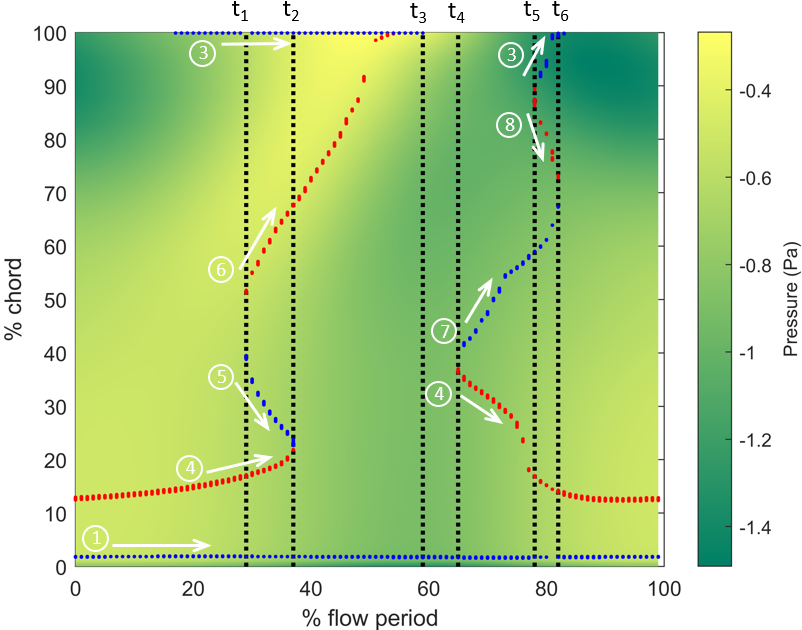}
}
\subfigure[Orthogonal selection array ($4 \times 10^3, 35^\circ$) \label{fig:physics_snap}]
{
\includegraphics[width=.48\linewidth, clip=true, trim=0in 0in 0in 0in]{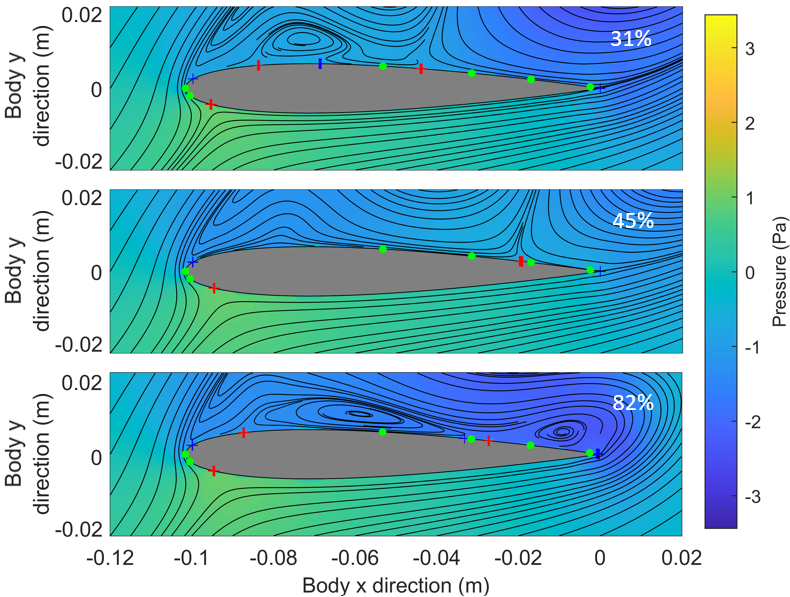}
}
\caption{Analysis of sensor arrays with respect to flow physics at operating point ($4 \times 10^3, 35^\circ$)}
\label{fig:physics}
\end{figure}

This section investigates how selected sensor locations correspond to physical features of the flow for the purpose of understanding how informative locations may change with changes in operating conditions. To examine the periodic behavior of the flow, the data from nine periodic shedding cycles were phase-averaged. Note that the phase averaging was performed with the initial flow transients removed from the data so that the flow has a consistent shedding frequency from cycle to cycle.  Figure \ref{fig:physics_st_pt} shows the evolution of the surface pressure on the suction side of the airfoil during the phase-averaged cycle of vortex shedding for the ($4 \times 10^3, 35^\circ$) operating point. The colormap displays the pressure along the upper surface of the airfoil ($y$-axis) during the phase-averaged period of vortex shedding ($x$-axis). In addition to the surface-pressure history, Fig.\ \ref{fig:physics_st_pt} displays markers denoting the chordwise locations of flow stagnation points (i.e., saddle points of the flow field) on the surface of the airfoil.

Stagnation point locations were identified by examining the flow velocity vectors at query points located $0.01c$ off of the airfoil's surface and searching for locations that maximized the magnitude of the dot product between the flow direction and the surface-normal vector from the closest point on the airfoil. Stagnation points lie on the wing near these locations of orthogonal flow. At each time instant, query points were selected as having flow nearly perpendicular to the surface if their normalized velocity vectors had a dot product greater than 0.9 with the surface normal vector. Only twenty locations were selected for any time instant in which more than twenty locations satisfied this criterion. Due to the close spacing of query points, several overlapping markers in Fig.\ \ref{fig:physics_st_pt} for a given time instant denote a single stagnation point. Stagnation points with flow velocity pointed towards (i.e., impinging on) the wing are colored with red markers, and stagnation points with flow directed away from the wing's surface are colored with blue markers. Gaps in the stagnation-point history in Fig.\ \ref{fig:physics_st_pt} suggest that this approach to stagnation-point identification can be further improved. Nevertheless, the method provides general trends for the movement of stagnation points on the upper surface of the airfoil that agree with the streamline topology of the flow field. Figure \ref{fig:physics_snap} shows snapshots of the flow at percentages of the shedding period indicated in the upper-right corner of each plot. The pressure field is plotted in the background of Fig.\ \ref{fig:physics_snap} with sensor locations from array E shown with green markers. Streamlines are shown in black, and `+' markers indicate stagnation points. The streamline topology in \ref{fig:physics_snap} motivates examination of informative sensor locations relative to important flow features, such as stagnation points.

\begin{figure}[t]
\centering
\includegraphics[width=.98\linewidth, clip=true, trim=0in 0in 0in 0in]{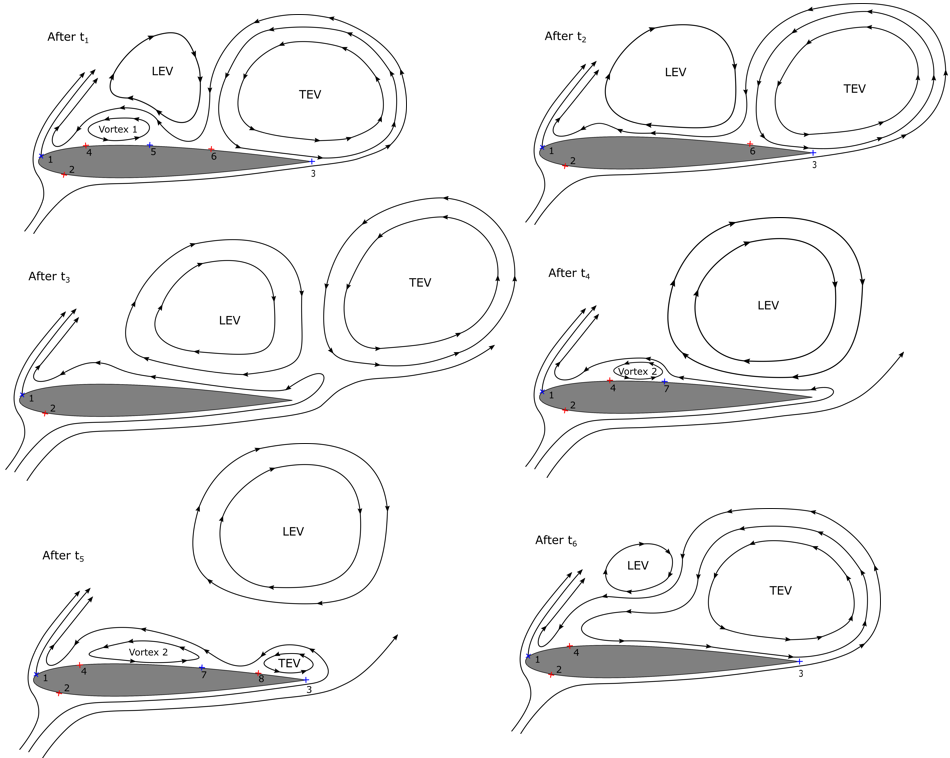}
\caption{Stagnation point locations and flow topology on the suction side of the wing for ($4 \times 10^3, 35^\circ$).}
\label{fig:topology}
\end{figure}

The markers in Fig.\ \ref{fig:physics_st_pt} show that the stagnation points move during the evolution of the vortex-shedding cycle. Note that the stagnation-point curves generally conform to the surface-pressure background in Fig.\ \ref{fig:physics_st_pt}, indicating that movement of the stagnation points is associated with changes in the surface pressure. An intersection of the red and blue markers at peaked or pointed locations of the curve in Fig.\ \ref{fig:physics_st_pt} signifies the appearance or disappearance of a pair of opposingly directed stagnation points, which signals a change in the flow topology. Six of these topological events occur, at the times $t_1$ through $t_6$ and are identified using vertical dotted lines in Fig.\ \ref{fig:physics_st_pt}. The general flow topology between these events is shown in Fig.\ \ref{fig:topology} using streamline illustrations. Note that these illustrations are hand drawn representations of the flow topology between topology enevts and are not drawn to scale. To aid in discussion, the stagnation points and vortices in Fig.\ \ref{fig:topology} are each assigned labels. Stagnation point 1 represents the separation point of flow near the leading edge. Stagnation point 2 marks where the flow impinges on the high-pressure side of the airfoil. Stagnation point 3 is at the trailing edge. Point 4 is an impinging stagnation point that lies between 10\% and 40\% chord throughout most of the shedding period. The remaining stagnation points have more movement due to topological events in the shedding period.

The first topological event occurs at $t_1$ when the Leading Edge Vortex (LEV) grows until it contacts the surface of the airfoil, creating stagnation points 5 and 6 as vortex 1 forms near the quarter chord. Note that vortex 1 is a portion of the Trailing Edge Vortex (TEV) that has been separated by contact of the LEV with the airfoil's surface. Vortex 1 steadily loses strength until it eventually diffuses away and stagnation points 4 and 5 coalesce and disappear at topological event $t_2$. Stagnation point 6 moves backward on the airfoil's surface until topological event $t_3$ marks shedding of the TEV. When the TEV sheds, saddle points 3 and 6 come together and stagnation point 6 becomes a saddle point that is shed into the flow. At $t_4$, vortex 2 forms due to the accelerated reverse flow caused by the growing of the LEV and creates stagnation points 4 and 7. Topological event $t_5$ marks the creation of the TEV and stagnation points 3 and 8. Shortly thereafter, the TEV grows in strength pushing stagnation points 7 and 8 together, which marks topological event $t_6$ and the shedding of the LEV. After $t_6$, vortex 2 merges into the TEV, and the periodic flow cycle continues again at $t_1$.

\begin{figure}[t]
\centering
\includegraphics[width=.98\linewidth, clip=true, trim=0in 0in 0in 0in]{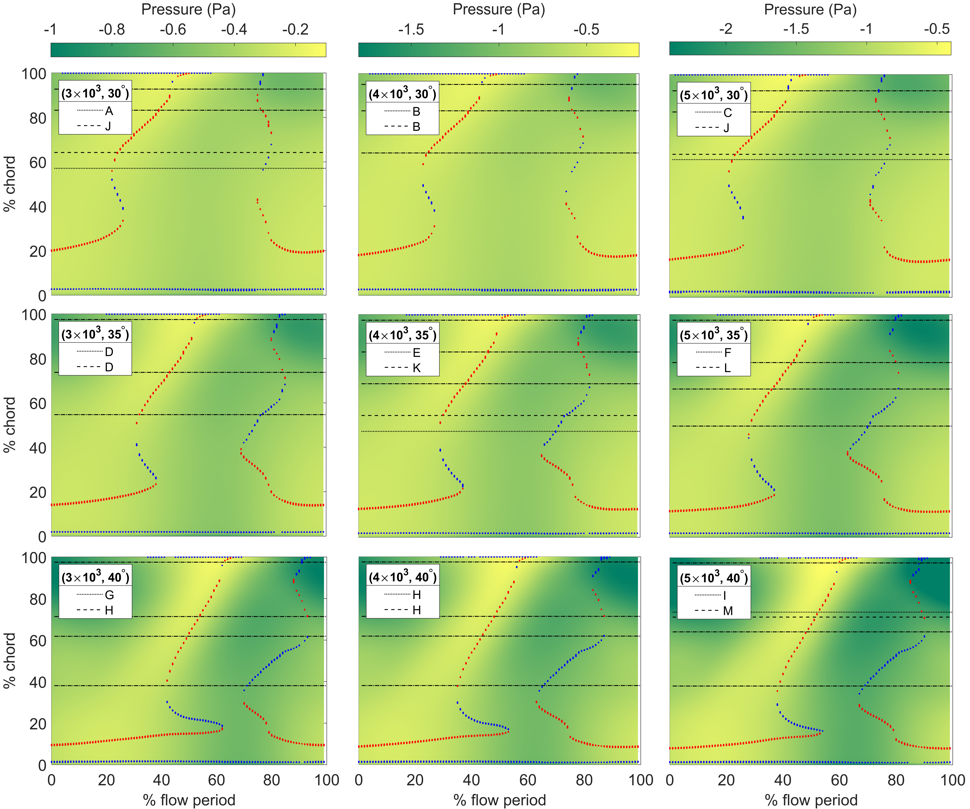}
\caption{Estimated stagnation point locations on the surface of the suction side of the wing}
\label{fig:st_pt}
\end{figure}

Figure \ref{fig:st_pt} shows the time evolution of the stagnation point locations for each of the nine operating points along with the locations of any sensors placed on the suction side of the airfoil for arrays designed at the indicated operating point. Horizontal dotted lines show the sensor locations for the array designed by orthogonal selection; dashed lines correspond to complementary selection. The stagnation points evolve with a similar structure for each of the nine operating points. Hence, Fig.\ \ref{fig:physics_st_pt} is representative of the general surface-pressure and stagnation-point dynamics that occur at all operating points considered in the paper. Changes in angle of attack cause the greatest differences in the stagnation-point curves, which is consistent with our observation regarding the information obtained by array designs in Sec. \ref{subsec:array_results}.

Each array designed in this paper (and shown in Fig.\ \ref{fig:all_arrays}) has one sensor placed on the leading edge and a second sensor placed near the leading edge on the high-pressure side of the airfoil.
These leading-edge sensors likely give insight into the speed of oncoming flow. Each array has a sensor placed near the trailing edge. Most of the arrays have the trailing-edge sensor on the suction side, except for arrays $A$, $B$, $C$, and $J$, that have this sensor on the high-pressure side of the airfoil. The trailing edge sensor likely identifies topological event $t_3$ (i.e. TEV shedding) since that event occurs at the trailing edge. Figure \ref{fig:st_pt} shows the locations of the remaining suction-side pressure sensors. In all of the arrays, the suction-side sensor that is closest to the leading edge is placed
just behind the location where the LEV contacts the airfoil in topological event $t_1$. More precisely, it appears to be placed near the appearance location of stagnation point 6, which separates the LEV from the TEV between $t_1$ and $t_3$ Knowledge of the existence and location of stagnation point 6 clearly provides information about the phase of the shedding cycle and the sizes and relative strengths of the LEV and TEV. The LEV and TEV are large, organizing structures, so determining their sizes and locations provides flow velocity information throughout the wake. Note that stagnation point 6 is also a desirable target for flow-sensing, because it can provide a strong pressure signal since it is an impinging stagnation point.

Additional sensors are spread out along the path taken by stagnation point 6 between $t_1$ and $t_3$. These sensors help track stagnation point 6, but they are also placed at locations that allows them to track other flow features. At the angles of attack of $35^\circ$ and $40^\circ$, sensors are placed near the location of topological event $t_6$, which sheds the LEV. At angle of attack of $30^\circ$, the sensors are placed near the location of TEV roll-up which occurs at topological event $t_5$. In total, the stagnation-point traces show that information-based sensor placement prefers locations near the occurrence of major topological events, such as the attachment and shedding locations of the largest vortices in the wake of the airfoil. Distribution of sensors along the path of an impinging stagnation point is also highly informative. 

%%%%%%%%%%%%%%%%%%%%%%%%%%%%%%%%%%%%%%%%%%%%%%%%%%%%%%%%%%%%%%%%%%%%%%%%%%%%%%%%%%%%%%%%%%%%%%%%%%%%%%%%%%%%%%%%%%%%

\section{Conclusion} \label{sec:Conclusion}

This paper presents a framework for information-based design of sensor arrays for use in data-driven estimation of unsteady flow fields. An intended application of this framework is flow estimation over an airfoil at various angles of attack using embedded pressure sensors.  The framework uses Kernel-based Dynamic Mode Decomposition to build a data-driven, linear model of the system that relates measurable output quantities to field quantities that are not measured. A Kalman filter estimates the flow state from the output measurements. The steady-state Kalman filter provides an information matrix that can be optimized in a resource-allocation problem that is independent of the flow's initial condition. To make the sensor selection sparse, two sequential techniques, orthogonal selection and complementary selection, are presented.  An example problem employs these methods for embedded pressure sensor placement using models built from simulated CFD data at nine different operating conditions.  Arrays created by the proposed orthogonal selection and complementary selection are shown perform comparable to or better than arrays built using QR pivoting sensor selection. To design a sensor array across several operating points, a composite system of all available flow models can be used during sensor selection.   The composite array performs well in filtering experiments across all operating points.  The composite measure of information for a given array evaluated at each possible operating condition gives a general prediction of filtering performance across operating conditions.  However, it was found that it was still possible to identify an array with better average filtering performance across operating conditions by selecting the array that maximizes the minimum information measure across all evaluation points.  Analysis of sensor placement results reveal that the placing pressure sensors near appearance location of an impinging stagnation point and along the path taken by the stagnation point during shedding is highly informative for the purpose of flow estimation.  Future research of the authors will examine the extension of the proposed sensor placement framework to three-dimensional flow fields.

\section*{Funding Sources}
This work was supported by the Air force Office of Scientific Research under award FA9550-21-1-0307.

\section*{Acknowledgements}
The authors gratefully acknowledge valuable discussions with Guillaume Sagnol and Takahito Isobe regarding this work.\\

\noindent Distribution Statement A: Approved for Public Release; Distribution is
Unlimited. PA\# AFRL-2023-1317.

\bibliography{references}

\begin{thebibliography}{31}
\newcommand{\enquote}[1]{``#1''}
\providecommand{\natexlab}[1]{#1}
\providecommand{\url}[1]{\texttt{#1}}
\providecommand{\urlprefix}{URL }
\expandafter\ifx\csname urlstyle\endcsname\relax
  \providecommand{\doi}[1]{\discretionary{}{}{}https://doi.org/#1}\else
  \providecommand{\doi}[1]{\discretionary{}{}{}\urlstyle{rm}\url{https://doi.org/#1}}\fi

\bibitem[{Tu et~al.(2013)Tu, Griffin, Hart, Rowley, Cattafesta, and
  Ukeiley}]{Tu_2013}
Tu, J.~H., Griffin, J., Hart, A., Rowley, C.~W., Cattafesta, L.~N., and
  Ukeiley, L.~S., \enquote{Integration of non-time-resolved PIV and
  time-resolved velocity point sensors for dynamic estimation of velocity
  fields,} \emph{Experiments in Fluids}, Vol.~54, No.~2, 2013, p. 1429.
\newblock \doi{10.1007/s00348-012-1429-7},
  \urlprefix\url{https://doi.org/10.1007/s00348-012-1429-7}.

\bibitem[{Gomez et~al.(2019)Gomez, Lagor, Kirk, Lind, Jones, and
  Paley}]{Gomez_2019}
Gomez, D.~F., Lagor, F.~D., Kirk, P.~B., Lind, A.~H., Jones, A.~R., and Paley,
  D.~A., \enquote{Data-Driven Estimation of the Unsteady Flowfield Near an
  Actuated Airfoil,} \emph{Journal of Guidance, Control, and Dynamics},
  Vol.~42, No.~10, 2019, pp. 2279--2287.
\newblock \doi{10.2514/1.G004339},
  \urlprefix\url{https://doi.org/10.2514/1.G004339}.

\bibitem[{Graff et~al.(2020)Graff, Ringuette, Singh, and Lagor}]{Graff_2019}
Graff, J., Ringuette, M.~J., Singh, T., and Lagor, F.~D.,
  \enquote{Reduced-Order Modeling for Dynamic Mode Decomposition without an
  Arbitrary Sparsity Parameter,} \emph{AIAA Journal}, Vol.~58, No.~9, 2020, pp.
  3919--3931.
\newblock \doi{10.2514/1.J059207}.

\bibitem[{Ahuja and Rowley(2010)}]{Ahuja_2010}
Ahuja, S., and Rowley, C.~W., \enquote{Feedback control of unstable steady
  states of flow past a flat plate using reduced-order estimators,}
  \emph{Journal of Fluid Mechanics}, Vol. 645, 2010, pp. 447--478.
\newblock \doi{10.1017/S0022112009992655},
  \urlprefix\url{https://doi.org/10.1017/S0022112009992655}.

\bibitem[{Rowley and Dawson(2017)}]{Rowley_Dawson_2017}
Rowley, C.~W., and Dawson, S.~T., \enquote{Model Reduction for Flow Analysis
  and Control,} \emph{Annual Review of Fluid Mechanics}, Vol.~49, No.~1, 2017,
  pp. 387--417.
\newblock \doi{10.1146/annurev-fluid-010816-060042},
  \urlprefix\url{https://doi.org/10.1146/annurev-fluid-010816-060042}.

\bibitem[{Joshi and Boyd(2009)}]{Joshi_2008}
Joshi, S., and Boyd, S., \enquote{Sensor Selection via Convex Optimization,}
  \emph{IEEE Transactions on Signal Processing}, Vol.~57, No.~2, 2009, pp.
  451--462.
\newblock \doi{10.1109/TSP.2008.2007095}.

\bibitem[{Shamaiah et~al.(2010)Shamaiah, Banerjee, and Vikalo}]{Shamaiah_2010}
Shamaiah, M., Banerjee, S., and Vikalo, H., \enquote{Greedy sensor selection:
  Leveraging submodularity,} \emph{49th IEEE Conference on Decision and Control
  (CDC)}, 2010, pp. 2572--2577.
\newblock \doi{10.1109/CDC.2010.5717225}.

\bibitem[{Hashemi et~al.(2021)Hashemi, Ghasemi, Vikalo, and
  Topcu}]{Hashemi_2020}
Hashemi, A., Ghasemi, M., Vikalo, H., and Topcu, U., \enquote{Randomized Greedy
  Sensor Selection: Leveraging Weak Submodularity,} \emph{IEEE Transactions on
  Automatic Control}, Vol.~66, No.~1, 2021, pp. 199--212.
\newblock \doi{10.1109/TAC.2020.2980924}.

\bibitem[{Willcox(2006)}]{WILLCOX2006}
Willcox, K., \enquote{Unsteady flow sensing and estimation via the gappy proper
  orthogonal decomposition,} \emph{Computers \& Fluids}, Vol.~35, No.~2, 2006,
  pp. 208--226.
\newblock \doi{https://doi.org/10.1016/j.compfluid.2004.11.006},
  \urlprefix\url{https://www.sciencedirect.com/science/article/pii/S0045793005000113}.

\bibitem[{Yang et~al.(2010)Yang, Venturi, Chen, Chryssostomidis, and
  Karniadakis}]{Yang_2010}
Yang, X., Venturi, D., Chen, C., Chryssostomidis, C., and Karniadakis, G.~E.,
  \enquote{EOF-based constrained sensor placement and field reconstruction from
  noisy ocean measurements: Application to Nantucket Sound,} \emph{Journal of
  Geophysical Research: Oceans}, Vol. 115, No. C12, 2010.
\newblock \doi{https://doi.org/10.1029/2010JC006148},
  \urlprefix\url{https://agupubs.onlinelibrary.wiley.com/doi/abs/10.1029/2010JC006148}.

\bibitem[{Manohar et~al.(2018)Manohar, Brunton, Kutz, and Brunton}]{QR_Manohar}
Manohar, K., Brunton, B.~W., Kutz, J.~N., and Brunton, S.~L.,
  \enquote{Data-Driven Sparse Sensor Placement for Reconstruction:
  Demonstrating the Benefits of Exploiting Known Patterns,} \emph{IEEE Control
  Systems Magazine}, Vol.~38, No.~3, 2018, pp. 63--86.
\newblock \doi{10.1109/MCS.2018.2810460}.

\bibitem[{Clark et~al.(2019)Clark, Askham, Brunton, and Kutz}]{QR_Clark}
Clark, E., Askham, T., Brunton, S.~L., and Kutz, J.~N., \enquote{Greedy Sensor
  Placement With Cost Constraints,} \emph{IEEE Sensors Journal}, Vol.~19,
  No.~7, 2019, pp. 2642--2656.
\newblock \doi{10.1109/JSEN.2018.2887044}.

\bibitem[{Saito et~al.(2021)Saito, Nonomura, Yamada, Nakai, Nagata, Asai,
  Sasaki, and Tsubakino}]{Saito_2021}
Saito, Y., Nonomura, T., Yamada, K., Nakai, K., Nagata, T., Asai, K., Sasaki,
  Y., and Tsubakino, D., \enquote{Determinant-Based Fast Greedy Sensor
  Selection Algorithm,} \emph{IEEE Access}, Vol.~9, 2021, pp. 68535--68551.
\newblock \doi{10.1109/ACCESS.2021.3076186}.

\bibitem[{Lagor et~al.(2013)Lagor, DeVries, Waychoff, and Paley}]{Lagor2013b}
Lagor, F.~D., DeVries, L.~D., Waychoff, K.~M., and Paley, D.~A.,
  \enquote{{Bio-inspired flow sensing and control: Autonomous rheotaxis using
  distributed pressure measurements},} \emph{Journal of Unmanned System
  Technology}, Vol.~1, No.~3, 2013, pp. 78--88.
\newblock \doi{10.21535\%2Fjust.v1i3.34}.

\bibitem[{Hinson and Morgansen(2014)}]{Hinson_2014}
Hinson, B.~T., and Morgansen, K.~A., \enquote{Observability-Based Optimal
  Sensor Placement for Flapping Airfoil Wake Estimation,} \emph{Journal of
  Guidance, Control, and Dynamics}, Vol.~37, No.~5, 2014, pp. 1477--1486.
\newblock \doi{10.2514/1.G000460},
  \urlprefix\url{https://doi.org/10.2514/1.G000460}.

\bibitem[{Lagor et~al.(2016)Lagor, Ide, and Paley}]{Lagor2016b}
Lagor, F.~D., Ide, K., and Paley, D.~A., \enquote{{Incorporating prior
  knowledge in observability-based path planning for ocean sampling},}
  \emph{Systems and Control Letters}, Vol.~97, 2016, pp. 169--175.
\newblock \doi{10.1016/j.sysconle.2016.09.002},
  \urlprefix\url{https://linkinghub.elsevier.com/retrieve/pii/S0167691116301189}.

\bibitem[{Bopardikar et~al.(2019)Bopardikar, Ennasr, and Tan}]{Bopardikar_2019}
Bopardikar, S.~D., Ennasr, O., and Tan, X., \enquote{Randomized Sensor
  Selection for Nonlinear Systems With Application to Target Localization,}
  \emph{IEEE Robotics and Automation Letters}, Vol.~4, No.~4, 2019, pp.
  3553--3560.
\newblock \doi{10.1109/LRA.2019.2928208}.

\bibitem[{Tzoumas et~al.(2016)Tzoumas, Jadbabaie, and Pappas}]{Tzoumas_2016}
Tzoumas, V., Jadbabaie, A., and Pappas, G.~J., \enquote{Sensor placement for
  optimal Kalman filtering: Fundamental limits, submodularity, and algorithms,}
  \emph{2016 American Control Conference (ACC)}, 2016, pp. 191--196.
\newblock \doi{10.1109/ACC.2016.7524914}.

\bibitem[{Zhang et~al.(2017)Zhang, Ayoub, and Sundaram}]{Zhang_2017}
Zhang, H., Ayoub, R., and Sundaram, S., \enquote{Sensor selection for Kalman
  filtering of linear dynamical systems: Complexity, limitations and greedy
  algorithms,} \emph{Automatica}, Vol.~78, 2017, pp. 202--210.
\newblock \doi{https://doi.org/10.1016/j.automatica.2016.12.025},
  \urlprefix\url{https://www.sciencedirect.com/science/article/pii/S0005109816305337}.

\bibitem[{Koopman(1931)}]{Koopman_1931}
Koopman, B.~O., \enquote{Hamiltonian Systems and Transformation in Hilbert
  Space,} \emph{Proceedings of the National Academy of Sciences}, Vol.~17,
  No.~5, 1931, pp. 315--318.
\newblock \doi{10.1073/pnas.17.5.315},
  \urlprefix\url{https://www.pnas.org/content/17/5/315}.

\bibitem[{Koopman and Neumann(1932)}]{Koopman_1932}
Koopman, B.~O., and Neumann, J.~V., \enquote{Dynamical Systems of Continuous
  Spectra,} \emph{Proceedings of the National Academy of Sciences}, Vol.~18,
  No.~3, 1932, pp. 255--263.
\newblock \doi{10.1073/pnas.18.3.255},
  \urlprefix\url{https://www.pnas.org/content/18/3/255}.

\bibitem[{Williams et~al.(2015)Williams, Rowley, and
  Kevrekidis}]{Williams_2015}
Williams, M.~O., Rowley, C.~W., and Kevrekidis, I.~G., \enquote{A kernel-based
  method for data-driven koopman spectral analysis,} \emph{Journal of
  Computational Dynamics}, Vol.~2, No.~2, 2015, pp. 247--265.

\bibitem[{Surana and Banaszuk(2016)}]{Surana_2016}
Surana, A., and Banaszuk, A., \enquote{Linear observer synthesis for nonlinear
  systems using Koopman Operator framework,} \emph{IFAC-PapersOnLine}, Vol.~49,
  No.~18, 2016, pp. 716 -- 723.
\newblock \doi{https://doi.org/10.1016/j.ifacol.2016.10.250}, 10th IFAC
  Symposium on Nonlinear Control Systems (NOLCOS), 2016.

\bibitem[{Sagnol and Harman(2015)}]{Sagnol_2015}
Sagnol, G., and Harman, R., \enquote{Optimal Designs for Steady-State Kalman
  Filters,} \emph{Stochastic Models, Statistics and Their Applications}, edited
  by A.~Steland, E.~Rafaj{\l}owicz, and K.~Szajowski, Springer International
  Publishing, Cham, 2015, pp. 149--157.

\bibitem[{Tu et~al.(2014)Tu, Rowley, Luchtenburg, Brunton, and Kutz}]{Tu2014}
Tu, J.~H., Rowley, C.~W., Luchtenburg, D.~M., Brunton, S.~L., and Kutz, J.~N.,
  \enquote{On Dynamic Mode Decomposition: Theory and applications.}
  \emph{Journal of Computational Dynamics}, Vol.~1, No.~2, 2014, pp. 391 --
  421.

\bibitem[{Simon(2006)}]{Simon_Book_CH5}
Simon, D., \emph{Optimal State Estimation}, John Wiley \& Sons, Ltd, 2006,
  Chap.~5, pp. 121--148.
\newblock \doi{https://doi.org/10.1002/0470045345.ch5}.

\bibitem[{Pukelsheim(2006)}]{Pukelsheim_2006}
Pukelsheim, F., \emph{Optimal Design of Experiments}, Society for Industrial
  and Applied Mathematics, 2006.
\newblock \doi{10.1137/1.9780898719109},
  \urlprefix\url{https://epubs.siam.org/doi/abs/10.1137/1.9780898719109}.

\bibitem[{Grant and Boyd(2014)}]{CVX1}
Grant, M., and Boyd, S., \enquote{{CVX}: Matlab Software for Disciplined Convex
  Programming, version 2.1,} \url{http://cvxr.com/cvx}, Mar. 2014.

\bibitem[{Grant and Boyd(2008)}]{CVX2}
Grant, M., and Boyd, S., \enquote{Graph implementations for nonsmooth convex
  programs,} \emph{Recent Advances in Learning and Control}, edited by
  V.~Blondel, S.~Boyd, and H.~Kimura, Lecture Notes in Control and Information
  Sciences, Springer-Verlag Limited, 2008, pp. 95--110.
\newblock \url{http://stanford.edu/~boyd/graph_dcp.html}.

\bibitem[{{COMSOL Inc}(2021)}]{COMSOL}
{COMSOL Inc}, \enquote{COMSOL Multiphysics Reference Manual, version 6.0,} ,
  2021.
\newblock \urlprefix\url{www.comsol.com}.

\bibitem[{Berkooz et~al.(1993)Berkooz, Holmes, and Lumley}]{Lumley_1993}
Berkooz, G., Holmes, P., and Lumley, J.~L., \enquote{The Proper Orthogonal
  Decomposition in the Analysis of Turbulent Flows,} \emph{Annual Review of
  Fluid Mechanics}, Vol.~25, No.~1, 1993, pp. 539--575.
\newblock \doi{10.1146/annurev.fl.25.010193.002543},
  \urlprefix\url{https://doi.org/10.1146/annurev.fl.25.010193.002543}.

\end{thebibliography}

\end{document}